\documentclass[onecolumn]{aastex701}  
\submitjournal{the Astrophysical Journal} 

\usepackage[english]{babel}
\usepackage{amssymb,amsmath,amsfonts,amsbsy}
\usepackage{aas_macros}
\usepackage{color}
\usepackage{enumerate}
\usepackage{graphicx}
\usepackage{pifont}
\usepackage{times}

\usepackage{subfigure}

\usepackage{cancel}

\usepackage{bm}

\usepackage[normalem]{ulem}

\begin{document}

\title{Non-Gaussian Expansion of Minkowski Tensors in Redshift Space}

\author{Stephen Appleby}
\email{stephen.appleby@apctp.org}
\affiliation{Asia Pacific Center for Theoretical Physics, Pohang 37673, Republic of Korea}
\affiliation{Department of Physics, POSTECH, Pohang 37673, Republic of Korea}
\author{Christophe Pichon}
\email{pichon@iap.fr} \correspondingauthor{Christophe Pichon}
\affiliation{Institut d'Astrophysique de Paris, 98 bis Boulevard Arago, F-75014 Paris, France}
\affiliation{
         Kyung Hee University, Dept. of Astronomy \& Space Science, Yongin-shi, Gyeonggi-do 17104, Republic of Korea }
\author{Pravabati Chingangbam}
\email{pravabati@gmail.com}
\affiliation{Indian Institute of Astrophysics, Koramangala II Block,  Bangalore  560 034, India}
\author{Dmitri Pogosyan}
\email{pogosyan@ualberta.ca}
\affiliation{Department of Physics, University of Alberta, 11322-89 Avenue, Edmonton, Alberta, T6G 2G7, Canada.}
\author{Changbom Park}
\email{cbp@kias.re.kr}
\affiliation{School of Physics, Korea Institute for Advanced Study, 85 Hoegiro, Dongdaemun-gu, Seoul, 02455, Korea}

\begin{abstract}
This paper focuses on extending the use of Minkowski Tensors to analyze anisotropic signals in cosmological data, focusing on those introduced by redshift space distortion. We derive the ensemble average of the two translation-invariant, rank-2 Minkowski Tensors ($W_1^{0,2}$ and $W_2^{0,2}$) for a matter density field that is perturbatively non-Gaussian in redshift space. This is achieved through the Edgeworth expansion of the joint probability density function of the field and its derivatives, expressing the ensemble averages in terms of cumulants up to cubic order. Our goal is to connect these theoretical predictions to the underlying cosmological parameters, allowing for parameter estimation by measuring them from galaxy surveys. The work builds on previous analyses of Minkowski Functionals in both real and redshift space and addresses the effects of Finger-of-God velocity dispersion and shot noise. We validate our predictions by matching them to measurements of the Minkowski Tensors from dark matter simulation data, finding that perturbation theory is a qualified success. Non-perturbative Finger-of-God effects remain significant at relatively large scales $R_\mathrm{G} \lesssim 20 \, h^{-1} \,  {\rm Mpc}$ and are particularly pronounced in the components parallel to the line of sight. 
\end{abstract}

\section{Introduction}
\label{sec:introduction}

As the Universe evolves, the non-linear nature of gravitational collapse generates non-Gaussianity in the matter distribution. Non-Gaussianity can be quantified via the $N$-point functions of cosmological fields, and by extracting these statistics from galaxy data we can better understand structure formation and infer cosmological parameters. A complicating factor in this scenario is that we almost always measure galaxy positions (velocities) in redshift space, which introduces anisotropy along the line of sight due to peculiar motion. The velocities of galaxies is not random; they are in-falling into  overdensities on all scales. Hence by measuring the distribution of galaxies in redshift space we are simultaneously observing large scale structure generated by gravitational collapse, and also the in-fall of galaxies into gravitational potentials generated by the structures. To extract information from anisotropic fields, statistical methodologies can be employed that are specifically designed to capture directional information.

One such example is the Minkowski tensors (MT), a rank-p generalization of the scalar Minkowski Functionals 
\citep{nla.cat-vn2176896,McMullen:1997,Alesker1999,2002LNP...600..238B,HugSchSch07,1367-2630-15-8-083028,JMI:JMI3331,Hadwiger},
which were first introduced to cosmology in \cite{Beisbart:2001vb,Beisbart:2001gk,2002LNP...600..238B,Ganesan:2017,Chingangbam:2017uqv}. While Minkowski functionals, frequently studied within the context of cosmology \citep{Gott:1989yj,1991ApJ...378..457P,Mecke:1994ax,Schmalzing:1997aj,Schmalzing:1997uc,1989ApJ...345..618M,1992ApJ...387....1P,2001ApJ...553...33P,Park:2009ja,doi:10.1111/j.1365-2966.2010.18015.x,Sahni:1998cr,Bharadwaj:1999jm,vandeWeygaert:2011ip, Park:2013dga,vandeWeygaert:2011hyr,Chingangbam:2013,Shivshankar:2015aza,Pranav:2016gwr,Pranav:2018lox,Pranav:2018pnu,Feldbrugge:2019tal,Wilding:2020oza,Munshi:2020tzm,Liu:2022vtr,Liu:2023qrj,Liu:2025haj,Chingangbam:2017PhLB,Rana:2018,Rahman:2021}, quantify scalar properties (volume, surface area, etc.), Minkowski tensors quantify directional information, making them ideal for analyzing anisotropic structures found in redshift space \citep{1987MNRAS.227....1K,1998ASSL..231..185H,1996MNRAS.282..877B,2011MNRAS.417.1913R,2007MNRAS.374..477T,2012JCAP...11..014O}. The Minkowski Tensors, and closely related statistics \citep{Kanafi:2023hmr,10.1093/mnras/staf1110}, are finding increasing application within cosmology \cite{Schmalzing:1997uc,Hikage:2012,Munshi:2013,Marques:2018ctl, Kapahtia:2018,Kapahtia:2019ksk,Goyal:2019vkq,Appleby:2020pem,Appleby:2021lfq,Appleby:2021xoz,K.:2018wpn,Kapahtia:2021,Goyal:2021,Liu:2024uxa,Chingangbam:2021,Bashir:2025}. Prior to the introduction of the Minkowski tensors, it was recognized that the Minkowski functionals extracted from two-dimensional subsets of redshift space data could be used to measure the growth rate $f$, by varying the angle between the two-dimensional slice and the line of sight \citep{1996ApJ...457...13M,Codis:2013exa}.

This paper presents a theoretical framework for the Minkowski Tensors extracted from a perturbatively non-Gaussian random field in redshift space.  We extend the work of \cite{2018ApJ...863..200A,Appleby_2019} to include the leading order cubic non-Gaussian contribution to the ensemble average of the Minkowski tensors. This is achieved via the Edgeworth expansion of the joint probability density function of the field and its first and second derivatives, and expressing the ensemble averages in terms of cumulants up to cubic order.  The work relies on previous analysis performed on the Minkowski functionals in real and redshift space \citep{Matsubara:1995dv,Matsubara:1995ns,1996ApJ...457...13M,2003ApJ...584....1M,Codis:2013exa,Gay:2011wz}, and uses these results to construct the corresponding ensemble averages of the tensors. The ensemble average of the Minkowski Functionals of random fields has been extensively studied \citep{1970Ap......6..320D,Adler,Gott:1986uz,10.1143/PTP.76.952,Hamilton:1986,Ryden:1988rk,1987ApJ...319....1G,1987ApJ...321....2W,Matsubara:2020fet,Matsubara:2020knr,Kuriki_Matsubara_2023,Matsubara:1994wn,Matsubara:1994we,1988ApJ...328...50M,
10.1111/j.1365-2966.2008.12944.x,Pogosyan:2009rg,Chingangbam:2024}, but the tensors to a far lesser degree. 

Our primary motivation is to express the theoretical predictions for the ensemble averages of the Minkowski Tensors to the underlying cosmological parameters using perturbation theory to describe the two- and three-point cumulants. This connection is useful because it provides a pathway for cosmological parameter estimation by measuring the tensors from observational data. The non-Gaussian components will  enhance the constraining power of the statistics by capturing additional information. We also include some complicating factors present in real observations, such as Finger-of-God velocity dispersion and shot noise. We validate our theoretical predictions by comparing them to measurements of the Minkowski tensors obtained from dark matter simulation data. The work serves as a precursor to measuring the Minkowski tensors from galaxy catalogs, so as to leverage cosmological parameters from the  corresponding cumulants.  

The rest of the paper will proceed as follows. In Section \ref{sec:rsd} we briefly introduce the plane parallel redshift space distorted field that we will study. In Section \ref{sec:definitions} we define the Minkowski tensors that are the main subject of the work, and in Section \ref{sec:ensav} construct their ensemble averages in terms of two- and three-point cumulants of the matter density field and its first and second derivatives. In Section \ref{sec:cumulants} we express the cumulants in terms of cosmology using perturbation theory results, and in Section \ref{sec:results} compare the ensemble averages to the numerical extraction of the statistics from dark matter snapshot data. We conclude in Section \ref{sec:disc}.

\section{Redshift Space Fields}
\label{sec:rsd}

In this work, we focus exclusively on plane parallel redshift space distorted fields and Cartesian tensors. In \cite{Appleby:2022itn} some of the authors considered the more complicated case of spherical redshift space distortion, in the Gaussian and Kaiserian limit. 

In real space, we assume that the matter density field $\delta$ is statistically isotropic and homogeneous. Throughout this work we also assume that the field is mean subtracted. The relation between the real (${\bf x}$) and redshift (${\bf s}$) space position of a particle is given by 
\begin{equation} 
{\bf s} = {\bf x} + {\bf e}_{3}  \, ({\bf v} \cdot{\bf e}_{3}) {(1+z) \over H(z)}, \end{equation} 
where ${\bf v}$ is the peculiar velocity and $H(z)$ is the Hubble parameter. A more accurate formula would include the effect of the observer peculiar velocity ${\bf v}_{0}$ via the replacement $\mathbf{v(x)} \to \mathbf{v(x)} - \mathbf{v_0}$, which accounts for the fact that the effect should decrease for closer objects over scales where peculiar velocities become correlated. In this work we fix ${\bf v}_{0} = 0$, assuming that the observer is sufficiently distant that velocity correlation is negligible.

The plane parallel approximation has already been implemented via the assumption every tracer particle is subject to a single, parallel line of sight, taken here to be ${\bf e}_{3}$. If we now assume a linear relation between the velocity and density field, then the density field in redshift space $\tilde{\delta}$ is related to the real space field $\delta$ according to 
\begin{equation}
\label{eq:pp1} \tilde{\delta}({\bf k}) = (1 + f \mu^{2}) \delta ({\bf k}) , \end{equation} 
 where $\mu = {\bf k}\cdot {\bf 
e}_{3}/|k|$ is the cosine of the angle between the line of sight and wavenumber ${\bf k}$, $f=d\ln D/d\ln a$ and $D$ is the linear growth factor. We note that due to the plane parallel approximation, the field $\tilde{\delta}({\bf k})$ is anisotropic but statistically homogeneous. The observed galaxy distribution is radially corrected, and the actual redshift space field is inhomogeneous due to the presence of a preferred location: the observer at ${\bf r}=0$.

The field defined in equation \eqref{eq:pp1} is the linear density field under the action of the linear Kaiser effect. The expectation value of the Minkowski tensors measured from such as field has been constructed in \cite{Appleby_2019}. The goal of this work is to calculate the ensemble average of the MTs for fields that are perturbatively non-Gaussian, in redshift space. There are two steps to performing this calculation. First, we use the Edgeworth expansion and the corresponding joint probability density function of the redshift space field and its first and second derivatives, and calculate the ensemble average of the Minkowski Tensors in terms of the cumulants. In this work, we only proceed to the three-point function (bispectrum) contributions, but one could continue to arbitrary order. Second, we write the cumulants in terms of the redshift space matter power and bi-spectra. The ultimate goal is to write the ensemble average in terms of cosmological parameters, which will allow us to perform parameter estimation by measuring the Minkowski tensors from cosmological datasets.

\section{Minkowski Tensors : Definitions}
\label{sec:definitions}
The Minkowski Tensors are rank-$p$ generalisations of the scalar Minkowski Functionals. To define these statistics, we begin with a smoothed mean zero density field $\delta(\mathbf{x})$ with variance $\sigma^{2} = \langle \delta^{2}
\rangle$, and define the excursion set at threshold $\nu$ as
\[
Q(\nu) = \{\, \mathbf{x} \; | \; \delta(\mathbf{x}) \ge \nu \sigma \,\}.
\]
This set contains all spatial regions where the density contrast exceeds the
chosen threshold, and its boundary $\partial Q$ is the two-dimensional
surface defined by $\delta = \nu \sigma$. For three-dimensional fields, two
vectors are associated with an excursion set $Q$: the position
vector $\mathbf{x}$ of a point within $Q$ (or on
$\partial Q$), and the unit vector $\hat{\mathbf{n}}$ normal to the boundary $\partial Q$. 
In this work we focus on rank-2 Minkowski Tensors, which are
the lowest-order generalisations of the scalar Minkowski Functionals that
have non-zero ensemble averages. The complete set of rank-2 tensors that can
be constructed from ${\bf x}$ and ${\bf \hat{n}}$ is \citep{1367-2630-15-8-083028}: 
\begin{eqnarray} & & W_{0}^{2,0} = \int_{Q} {\bf x}^{2} \textrm{dV} \, , \\
& & W_{k}^{r, s} = \int_{\partial Q} G_{k} {\bf x}^{r} {\bf \hat{n}}^{s} \textrm{dA}  \, ,
\end{eqnarray} 
where $r+s = 2$, and ${\bf x}^{r} = {\bf x} \otimes {\bf x} \dots$, ${\bf \hat{n}}^{s} = {\bf \hat{n}} \otimes {\bf \hat{n}} \dots$ are symmetric tensor products of ${\bf x}$ and ${\bf \hat{n}}$, $r$ and $s$ times, respectively. The factors $G_k$ are: $G_{1} = 1$, $G_{2}$ is the mean curvature of the surface $\partial Q$, and $G_{3}$ its Gaussian curvature. $\textrm{dV}$ and $\textrm{dA} $ are the volume and area elements, respectively.
Of these tensors, we restrict our analysis to the ones  that are translation invariant, which are $W_{1,2,3}^{0,2}$, $W_{1,2,3}^{1,1}$. Between these quantities, linear relations exist of the form \citep{1367-2630-15-8-083028};
\begin{equation} {\mathbb I} W_k = k W_k^{0,2} + (3-k) W_{k+1}^{1,1},
\end{equation} 
where ${\mathbb I}$ is the identity matrix, $k = 0,1,2,3$ and $W_k$ are the scalar MFs. These relations imply that $W_{3}^{0,2}$ contains no additional information relative to $W_{3}$, and only two of $W_{1,2}^{0,2}$, $W_{2,3}^{1,1}$ need to be studied beyond the scalar Minkowski Functionals. Of these, we focus on  
\begin{subequations} \label{eq:defWMT}
\begin{eqnarray} & & W_{1}^{0,2} = {1 \over 6V} \int_{\partial Q}\! \hat{n} \otimes \hat{n}\, \textrm{dA} \,, \\ 
& & W_{2}^{0,2} = {1 \over 3\pi V} \int_{\partial Q}\! \hat{n} \otimes \hat{n} \,  G_{2}  \textrm{dA} \, .
\end{eqnarray} 
\end{subequations}
Note that in the above equations we have normalized by appropriate volume factors, $V$ being the volume over which the field $\delta$ is defined. The mean curvature $G_{2}$ can be written as 
\begin{eqnarray}\nonumber G_{2} &=& -{1 \over 2} \nabla \cdot {\bf \hat{n}} = -{1 \over 2} \left[ {\nabla^{2}\delta \over |\nabla \delta|} - {\nabla \delta \over |\nabla \delta|^{2}} \cdot \nabla |\nabla \delta|  \right] \, , \\
\nonumber &=& -{1 \over 2 |\nabla \delta|^{3}} \left[ (\delta_{2}^{2}+\delta_{3}^{2}) \delta_{11} + (\delta_{1}^{2}+\delta_{3}^{2}) \delta_{22} + (\delta_{1}^{2}+\delta_{2}^{2}) \delta_{33}    \right]  \\ 
\label{eq:G2} & & +{1 \over  |\nabla \delta|^{3}} \left[ \delta_{1}\delta_{2}\delta_{12} + \delta_{1}\delta_{3}\delta_{13} + \delta_{2}\delta_{3}\delta_{23}  \right]\, ,
\end{eqnarray}
where $\delta_{i} = \nabla_{i} \delta$. We can also write the unit normal in terms of derivatives of the field $\hat{n}_{i} = \delta_{i}/|\nabla \delta|$. $W_{1}^{0,2}$ and $W_{2}^{0,2}$ represent tensorial generalisations of the scalar Minkowski Functionals $W_{1}$ and $W_{2}$. They describe the surface integrals of the excursion set boundary, now weighted by the product of normals to the surface at each point.

 \citep{Schmalzing:1997aj,Schmalzing:1997uc} have shown how to  re-write the MTs \eqref{eq:defWMT} as volume integrals
\begin{subequations}
\begin{eqnarray} & & W_{1}^{0,2}|_{i}{}^{j} = {1 \over 6V} \int \textrm{dV} \delta_\mathrm{D}(\delta - \nu\sigma ) {\delta_{i}  \delta^{j} \over |\nabla \delta|}\,, \\
& & W_{2}^{0,2}|_{i}{}^{j} = {1 \over 3\pi V} \int \textrm{dV} \delta_\mathrm{D}(\delta - \nu\sigma ) \, G_{2} {\delta_{i} \delta^{j} \over |\nabla \delta|}\,, 
\end{eqnarray} 
\end{subequations}
where we have introduced $\delta_\mathrm{D}(x)$ as the delta function. The  $ W^{0,2}|_{i}{}^{j}$ are respectively the volume averages of the quantities 
\begin{subequations}
\begin{eqnarray}
& & w_{i}{}^{j} \equiv {1 \over 6} \delta_\mathrm{D}(\delta - \nu\sigma) {\delta_{i}\delta^{j} \over \sqrt{\delta_{k}\delta^{k}}} \, , \\
& & v_{i}{}^{j} \equiv {1 \over 3\pi} \delta_\mathrm{D}(\delta - \nu\sigma) G_{2} {\delta_{i}\delta^{j} \over \sqrt{\delta_{k}\delta^{k}}} \, .
\end{eqnarray} 
\end{subequations}
In the following section, we construct the ensemble averages $\langle w_{i}{}^{j} \rangle$ and $\langle v_{i}{}^{j} \rangle$ using the Edgeworth expansion. The former is only a function of the field $\delta$ and its first derivatives, whereas the latter is also a function of the second derivatives of the field due to the $G_{2}$ term.\footnote{In this work we focus on calculating the components of the tensors in a particular coordinate system, and we direct the reader to \cite{Matsubara:2022ohx,Matsubara:2022eui,Matsubara:2023avg,Matsubara:2024sqn} for a more general formalism for extracting information from tensors.}

\section{Ensemble Averages}
\label{sec:ensav}
%
We start by defining the ensemble averages 
\begin{equation}  
\langle w_{i}{}^{j} \rangle = \int w_{i}{}^{j} P(X) dX \, , 
\qquad  \langle v_{i}{}^{j} \rangle = \int v_{i}{}^{j} P(X') dX' \, ,
\end{equation} 
where $X=(\delta, \delta_{i})$ and $X' = (\delta, \delta_{i}, \delta_{jk})$ are the set of random variables to which $w_{i}{}^{j}$, $v_{i}{}^{j}$ are sensitive and $P(X)$, $P(X')$ their joint probability density functions. By utilising linearity of the ensemble average (it is effectively a weighted sum of states) and the assumed statistical homogeneity of the field, the volume and ensemble averages in $\langle W_{1}^{0,2}|_{i}{}^{j}\rangle$, $\langle W_{2}^{0,2}|_{i}{}^{j}\rangle$ commute, so we can write $\langle w_{i}{}^{j} \rangle = \langle W_{1}^{0,2}|_{i}{}^{j}\rangle$, $\langle v_{i}{}^{j} \rangle = \langle W_{2}^{0,2}|_{i}{}^{j}\rangle$. Here we make the implicit assumption that the volume averages constructed from data are representative of the integral over all field configurations which defines the ensemble average. In this work we do not study the uncertainty or potential biases that may be arise from the finite volume of our simulations, we simply assume ergodicity holds exactly.

The integrals over the random fields can be simplified if the fields exhibit some symmetry. For example, in real space $\delta({\bf x})$ is isotropic and we can use $\langle \delta_{j}\delta_{k} \rangle \propto {\mathbb I}_{jk}$, where ${\mathbb I}$ is the identity matrix. In redshift space, rotational invariance is partially broken along the line of sight, but the field is assumed to be statistically isotropic in the two-dimensional subspace perpendicular to this direction. We expect the Minkowski tensors to respect this symmetry, and be of the form $\propto {\rm diag}[a,a,b]$ in a Cartesian coordinate system with basis ${\bf e}_{1}, {\bf e}_{2}, {\bf e}_{3}$ that is aligned with the line of sight ${\bf e}_{3}$. A different choice of coordinates that is not aligned with ${\bf e}_{3}$ would generate off-diagonal components. Changes in coordinate systems can be represented by rotation matrices operating on the tensors, and only for an isotropic field are $\langle w_{i}{}^{j} \rangle$ and $\langle v_{i}{}^{j} \rangle$ diagonal in every coordinate system. Note that we could instead study coordinate invariant quantities such as the eigenvalues of $w_{i}{}^{j}$ and $v_{i}{}^{j}$, as originally proposed in \cite{Chingangbam:2017uqv}. However, these are non-linear functions of the tensor elements and their ensemble averages are not currently known.

\subsection{Ensemble average of $w_{i}{}^{j}$}
\label{sec:ens_w}

We follow \cite{Codis:2013exa} closely throughout, and so adopt the same variables as that work; $x=\delta/\sigma$, $q_{\perp}^{2} = (\delta_{1}^{2} + \delta_{2}^{2})/\sigma_{1\perp}^{2}$, $x_{3} = \delta_{3}/\sigma_{1\parallel}$. The cumulants $\sigma^{2}$, $\sigma_{1\parallel}^{2}$ and $\sigma_{1\perp}^{2}$ are the field variances for $\delta$, $\delta_{3}$ and $\sqrt{\delta_{1}^{2} + \delta_{2}^{2}}$ respectively. The joint probability density function (PDF) $P(x,q_{\perp}^{2},x_{3})$, up to cubic order in the cumulants, is given by 
\begin{eqnarray} 
P(x,q_{\perp}^{2},x_{3}) =  {1 \over 2\pi} e^{-x^{2}/2 - q_{\perp}^{2} - x_{3}^{2}/2}  \left[ 1 + \sum_{\kappa_{3}} {(-1)^{j} \over i! j! k!} \langle x^{i}q_{\perp}^{2j}x_{3}^{k}\rangle  \, H_{i}(x)L_{j}(q_{\perp}^{2}) H_{k}(x_{3}) \right]\,,
\end{eqnarray}
where the sum over $\kappa_{3} = \{i+2j+k = 3 \}$ is over all combinations of integers $i,2j,k \geq 0$ that sum to three. $H_{i}(x)$ and $L_{k}(x)$ are {\bf probabilist} Hermite and Laguerre polynomials respectively\footnote{The first few of which are $H_{0}=1$, $H_{1}=x$, $H_{2} = x^{2}-1$, $H_{3} = x^{3}-3x$ and $L_{0}=1$, $L_{2} = 1-x$, $L_{3} = (x^{2} - 4x + 2)/2$.}.

Using these dimensionless variables, the MT can be written as \begin{subequations}
\begin{eqnarray} & & w_{1}{}^{1}+w_{2}{}^{2} =  {\sigma_{1\perp}\over 6\sigma} \delta_\mathrm{D}(x - \nu) {q_{\perp}^{2} \over \tilde{X}}\,, \\
& & w_{3}{}^{3} =  {\sigma_{1\perp}\lambda^{2}\over 6\sigma} \delta_\mathrm{D}(x - \nu) {x_{3}^{2} \over \tilde{X}}\,, 
\end{eqnarray} 
\end{subequations}
 where $\tilde{X} =  \sqrt{q_{\perp}^{2}+\lambda^{2}x_{3}^{2}}$ and $\lambda = \sigma_{1\parallel}/\sigma_{1\perp}$. All off-diagonal terms will be zero in this coordinate system. 
 The ensemble average of the non-zero components of $w_{i}{}^{j}$ are then given by 
\begin{subequations}
    \begin{eqnarray} & & \langle w_{1}{}^{1} \rangle + \langle w_{2}{}^{2} \rangle = {\sigma_{1\perp} \over 6\sigma} \int  dq_{\perp}^{2}  dx_{3} P(\nu, q_{\perp}^{2}, x_{3})  {q_{\perp}^{2} \over \tilde{X}} \, , \\ 
& & \langle w_{3}{}^{3} \rangle  = {\sigma_{1\perp} \over 6\sigma} \int  dq_{\perp}^{2}  dx_{3} P(\nu, q_{\perp}^{2}, x_{3})  {\lambda^{2} x_{3}^{2} \over \tilde{X}} \, ,
\end{eqnarray} 
\end{subequations}
 where the integrals are over $x_{3} \in (-\infty,\infty)$ and $q_{\perp}^{2} \in [0, \infty)$. Integrating over the gradients $q_{\perp}^{2}, x_{3}$ using \verb"Mathematica", we find 
\begin{subequations}
\label{eq:w1}
\begin{eqnarray} \label{eq:w11} \langle w_{1}{}^{1} \rangle + \langle w_{2}{}^{2} \rangle  &=& {\sigma_{1\perp} e^{-\nu^{2} /2}\over 6\pi \sigma  }\left[ A^{(1)}_{\rm G \perp}\left(H_{0}(\nu) + {\langle x^{3} \rangle \over 6} H_{3}(\nu)\right) + \bigg( B^{(1)}_{\perp}\langle x q_{\perp}^{2} \rangle  + C^{(1)}_{\perp} \langle x x_{3}^{2} \rangle   \bigg) H_{1}(\nu) + {\cal O}(\sigma^{2})  \right] \,,\\
\label{eq:w33} \langle w_{3}{}^{3} \rangle &=&   {\sigma_{1\perp} e^{-\nu^{2} /2}\over 6\pi \sigma  }\left[ A^{(1)}_{\rm G \parallel}\left(H_{0}(\nu) + {\langle x^{3} \rangle \over 6} H_{3}(\nu)\right) + \bigg( B^{(1)}_{\parallel}\langle x q_{\perp}^{2} \rangle  + C^{(1)}_{\parallel} \langle x x_{3}^{2} \rangle   \bigg) H_{1}(\nu) +  {\cal O}(\sigma^{2}) \right]\,, 
\end{eqnarray} 
\end{subequations}
where the coefficients depend on $\lambda$ and are given by 
\begin{subequations}
    \begin{eqnarray} & &  A^{(1)}_\mathrm{G\perp} =  {\sqrt{2}(4\lambda^{2}-1) \over 4 (2\lambda^{2}-1)^{3/2}} \cosh^{-1}(4\lambda^{2}-1) - {\lambda \over 2\lambda^{2}-1}  \, , \\
& & B^{(1)}_{\perp} =  {\lambda (10\lambda^{2} - 16\lambda^{4} - 1) \over 2(2\lambda^{2}-1)^{3}} + {16\lambda^{4} - 4\lambda^{2} + 1  \over 4\sqrt{2}(2\lambda^{2}-1)^{5/2}}\cosh^{-1}(4\lambda^{2}-1) \,,\\
& & C^{(1)}_{\perp} =  {3\lambda^{3} \over (2\lambda^{2}-1)^{2}} - {\sqrt{2}(8\lambda^{2}-1) \over 4 (2\lambda^{2}-1)^{5/2}}\cosh^{-1}(\sqrt{2}\lambda) - {\sqrt{2}(4\lambda^{2}-1) \over 8 (2\lambda^{2}-1)^{3/2}}\cosh^{-1}(4\lambda^{2}-1)   \,,\\
& & A^{(1)}_\mathrm{G\parallel} =  {2\lambda^{3} \over 2\lambda^{2} - 1} - {\sqrt{2}\lambda^{2}  \over (2\lambda^{2} - 1)^{3/2}} \cosh^{-1}(\sqrt{2}\lambda)   \,, \\
& & B^{(1)}_{\parallel} = {3\lambda^{3} \over (2\lambda^{2}-1)^{2}} - {\lambda^{2}(1+4\lambda^{2}) \over \sqrt{2}(2\lambda^{2}-1)^{5/2}}\cosh^{-1}(\sqrt{2}\lambda) \,,\\
& & C^{(1)}_{\parallel} = {2\lambda^{3} (2\lambda^{4}-5\lambda^{2} + 2) \over (2\lambda^{2}-1)^{3}} + {\sqrt{2}\lambda^{2}(1+\lambda^{2}) \over 2(2\lambda^{2}-1)^{5/2}} \cosh^{-1}(4\lambda^{2}-1) \,. 
\end{eqnarray}
\end{subequations}

The Gaussian limit of the ensemble average is 
\begin{subequations}\label{eq:w1g}
\begin{eqnarray}  \langle w_{1}{}^{1} \rangle_\mathrm{G} + \langle w_{2}{}^{2} \rangle_\mathrm{G}  &=& {\sigma_{1\perp} e^{-\nu^{2} /2}\over 6\pi \sigma  } A^{(1)}_{\rm G \perp} H_{0}(\nu)  \,,\\
 \langle w_{3}{}^{3} \rangle_\mathrm{G} &=&   {\sigma_{1\perp} e^{-\nu^{2} /2}\over 6\pi \sigma  } A^{(1)}_{\rm G \parallel}H_{0}(\nu)  \,, 
\end{eqnarray} 
\end{subequations}
The trace of the ensemble averages (\ref{eq:w1g}) and (\ref{eq:w1}) should respectively yield the expectation value of the scalar Minkowski Functional $W_{1}$ in redshift space for a Gaussian, Kaiser corrected field and the leading order non-Gaussian result derived in \cite{Matsubara:1995wj,Codis:2013exa}. We have verified numerically that our results agree with the previous literature, in particular we present a comparison between our results and those of \citet{Codis:2013exa} in the Appendix (cf. Figure \ref{fig:app0}).

\subsection{Ensemble average of $v_{i}{}^{j}$}  
\label{sec:ens_v}

The ensemble average of $v_{i}{}^{j}$ proceeds similarly to $w_{i}{}^{j}$, but with some complicating factors. We again define everything in terms of the dimensionless variables introduced in \cite{Codis:2013exa}. In addition to $x$,  $q_{\perp}^{2}$ and $x_{3}$ used in the previous section, we introduce the terms $x_{1} = \delta_{1}/\sigma_{1\perp}$, $x_{2} = \delta_{2}/\sigma_{1\perp}$, $J_{1\perp} = (\delta_{11} + \delta_{22})/\sigma_{2\perp}$, $x_{11} = \delta_{11}/\sigma_{2\perp}$, $x_{22} = \delta_{22}/\sigma_{2\perp}$,  $x_{33} = \delta_{33}/\sigma_{2\parallel}$, $x_{12} =  \delta_{12}/\sigma_{2\perp}$, $x_{13} = \delta_{13}/\sigma_{Q}^{2}$ and $x_{23} = \delta_{23}/\sigma_{Q}^{2}$. These quantities have variances 
\begin{equation}
\langle x_{1}^{2} \rangle = \langle x_{2}^{2} \rangle = {1 \over 2} \, ,  \quad \langle J_{1\perp}^{2} \rangle = 1 \, ,  \quad \langle x_{11}^{2} \rangle =  \langle x_{22}^{2} \rangle = {3 \over 8} \, , \quad  \langle x_{3}^{2} \rangle = 1 \, ,  \quad  \langle x_{12}^{2} \rangle = {1 \over 8} \, , \quad \langle x_{13}^{2} \rangle =  \langle x_{23}^{2} \rangle = {1 \over 2} \, .
\end{equation} 
In terms of these variables, we have 
\begin{subequations}
    \begin{eqnarray} & & v_{1}{}^{1}+v_{2}{}^{2} =  {\sigma_{1\perp}\over 3\pi\sigma} \delta_\mathrm{D}(x - \nu) G_{2}{q_{\perp}^{2} \over \tilde{X}}\,, \\
& & v_{3}{}^{3} =  {\sigma_{1\perp}\lambda^{2}\over 3\pi\sigma} \delta_\mathrm{D}(x - \nu) G_{2} {x_{3}^{2} \over \tilde{X}}\,, 
\end{eqnarray} 
\end{subequations}
with  \begin{subequations}
    \begin{eqnarray}\nonumber G_{2} &=& -{1 \over 2 \sigma_{1\perp} \tilde{X}^{3}} \left[ (x_{2}^{2}+\lambda^{2}x_{3}^{2})  \sigma_{2\perp}  x_{11} + (x_{1}^{2} + \lambda^{2} x_{3}^{2}) \sigma_{2\perp}  x_{22} + \sigma_{2\parallel} q_{\perp}^{2}x_{33}    \right]  \\ 
\label{eq:G2b} & & +{1 \over  \sigma_{1\perp}\tilde{X}^{3}} \left[ \sigma_{2\perp} x_{1}x_{2} x_{12} + \lambda \sigma_{Q}  x_{1}x_{3}x_{13} + \lambda \sigma_{Q} x_{2}x_{3}x_{23}  \right]\, .
\end{eqnarray}
\end{subequations}
We note that there are six terms in $G_{2}$, and each contains a single second derivative. We can calculate the ensemble average of each separately. The terms proportional to $x_{11}$, $x_{22}$ and $x_{33}$ must be de-correlated from $x$ using the new variables 
\begin{equation}
y_{11} = {x_{11} + \gamma_{\perp}x \over \sqrt{1 - \gamma_{\perp}^{2}}} \,,
\quad
y_{22} = {x_{22} + \gamma_{\perp}x \over \sqrt{1 - \gamma_{\perp}^{2}}} \,,
\quad
y_{33} = {x_{33} + \gamma_{\parallel}x \over \sqrt{1 - \gamma_{\parallel}^{2}}} \,,
\end{equation}
 where $\gamma_{\perp} = \sigma_{1\perp}^{2}/(\sigma\sigma_{2\perp})$ and $\gamma_{\parallel} = \sigma_{1\parallel}^{2}/(\sigma\sigma_{2\parallel})$. We do not need to de-correlate the second derivatives from one another using the more complicated transformation used in \cite{Codis:2013exa}, because all second derivatives appear linearly in $v_{i}{}^{j}$, meaning that all components except one can be marginalised out for each term in $G_{2}$. Finally, each of the six contributions to $\langle v_{i}{}^{j}\rangle$ can be integrated over the joint probability $P(x,x_{1},x_{2},x_{3},\kappa)$, where $\kappa = (x_{11},x_{22}, x_{33},x_{12}, x_{13}, x_{23})$ depending on the term being calculated. The final integrals can be transformed from $(x_{1},x_{2},x_{3})$ to $(q_{\perp}^{2},x_{3})$ using the symmetry of the field, and we can use $\langle x_{i}^{2} x_{ii} \rangle = 0$, $\langle x_{1}^{2}x_{33} \rangle / \gamma_{\parallel} = \langle x_{3}^{2}x_{11} \rangle /\gamma_{\perp}$ to simplify the final result. After a tedious calculation, we can write the ensemble averages purely in terms of the variables defined in \cite{Codis:2013exa} as 
\begin{subequations}
\label{eq:v11}
\begin{eqnarray} \nonumber  \langle v_{1}{}^{1} \rangle + \langle v_{2}{}^{2} \rangle  &=& {\sigma_{1\perp}^{2} \over 12\pi\sqrt{2\pi}\sigma^{2}} e^{-\nu^{2}/2}\left[ A^{(2)}_{\rm G \perp}\left(H_{1}(\nu) + {\langle x^{3} \rangle \over 6} H_{4}(\nu)\right) + \bigg( B^{(2)}_{\perp}\langle x q_{\perp}^{2} \rangle  + C^{(2)}_{\perp} \langle x x_{3}^{2} \rangle   \bigg) H_{2}(\nu)  \right. \\
\label{eq:v1122} & &  + \bigg( D^{(2)}_{\perp}{\langle q_{\perp}^{2} J_{1\perp}  \rangle \over \gamma_{\perp}}  + E^{(2)}_{\perp} {\langle  x_{3}^{2} J_{1\perp} \rangle \over \gamma_{\perp}}    \bigg) H_{0}(\nu)   + {\cal O}(\sigma^{2})  \biggr] \,,\\ 
\nonumber \langle v_{3}{}^{3} \rangle &=&   {\sigma_{1\perp}^{2} \over 12\pi\sqrt{2\pi}\sigma^{2}} e^{-\nu^{2}/2}\left[ A^{(2)}_{\rm G \parallel}\left(H_{1}(\nu) + {\langle x^{3} \rangle \over 6} H_{4}(\nu)\right) + \bigg( B^{(2)}_{\parallel}\langle x q_{\perp}^{2} \rangle  + C^{(2)}_{\parallel} \langle x x_{3}^{2} \rangle   \bigg) H_{2}(\nu)   \right. \\
\label{eq:v33} & &  + \bigg( D^{(2)}_{\parallel}{\langle q_{\perp}^{2} J_{1\perp}  \rangle \over \gamma_{\perp}}  + E^{(2)}_{\parallel} {\langle  x_{3}^{2} J_{1\perp} \rangle \over \gamma_{\perp}}   \bigg) H_{0}(\nu)   + {\cal O}(\sigma^{2})  \biggr]\,,
\end{eqnarray} 
\end{subequations}
where 
\begin{subequations}
    \begin{eqnarray} & &  A^{(2)}_\mathrm{G\perp} =   {2\lambda^{4} - 3\lambda^{2} + 1 + 2\lambda^{4}\sqrt{2\lambda^{2}-1}\tan^{-1}\sqrt{2\lambda^{2}-1} \over (2\lambda^{2}-1)^{2} }   \, , \\
& & B^{(2)}_{\perp} = {2\lambda^{6} - 9\lambda^{4} + 6\lambda^{2} - 1 + 2\lambda^{4}(1+\lambda^{2})\sqrt{2\lambda^{2}-1}\tan^{-1}\sqrt{2\lambda^{2}-1} \over (2\lambda^{2}-1)^{3} }  \,,\\
& & C^{(2)}_{\perp} = {2\lambda^{6} + \lambda^{4} - \lambda^{2}  + 2\lambda^{4}(\lambda^{2}-2)\sqrt{2\lambda^{2}-1}\tan^{-1}\sqrt{2\lambda^{2}-1} \over (2\lambda^{2}-1)^{3} }  \,,\\
& & D^{(2)}_{\perp} = {10\lambda^{4} - 7\lambda^{2} + 1 - 6\lambda^{4}\sqrt{2\lambda^{2}-1} \tan^{-1}\sqrt{2\lambda^{2}-1} \over (2\lambda^{2} -1 )^{3}} \,,\\
& & E^{(2)}_{\perp} = {-2\lambda^{6} - \lambda^{4} + \lambda^{2} - 2\lambda^{4}(\lambda^{2}-2)\sqrt{2\lambda^{2}-1}\tan^{-1}\sqrt{2\lambda^{2}-1} \over (2\lambda^{2}-1)^{3}}  \,,\\
& & A^{(2)}_\mathrm{G\parallel} =  {2\lambda^{4} - \lambda^{2}  + 2\lambda^{2}(\lambda^{2} - 1)\sqrt{2\lambda^{2}-1} \tan^{-1}\sqrt{2\lambda^{2}-1}  \over (2\lambda^{2}-1)^{2} }    \,, \\
& & B^{(2)}_{\parallel} = {2\lambda^{6} + \lambda^{4} - \lambda^{2} + 2 \lambda^{4}(\lambda^{2}-2)\sqrt{2\lambda^{2}-1}\tan^{-1}\sqrt{2\lambda^{2}-1} \over  (2\lambda^{2}-1)^{3}} \,,\\
& & C^{(2)}_{\parallel} = {2\lambda^{6} - 5\lambda^{4} + 2\lambda^{2}  + 2(\lambda^{6} - \lambda^{4} + \lambda^{2})\sqrt{2\lambda^{2}-1}\tan^{-1}\sqrt{2\lambda^{2}-1} \over (2\lambda^{2}-1)^{3} }  \,, \\
& & D^{(2)}_{\parallel} = {-6\lambda^{4} + 3\lambda^{2} + 2\lambda^{2}(\lambda^{2}+1)\sqrt{2\lambda^{2}-1} \tan^{-1}\sqrt{2\lambda^{2}-1} \over  (\lambda^{2}-1)^{3}}  \,,\\
& & E^{(2)}_{\parallel} = {-2\lambda^{6} + 5\lambda^{4} - 2\lambda^{2} -2\lambda^{2}(\lambda^{4} - \lambda^{2} + 1) \sqrt{2\lambda^{2}-1} \tan^{-1} \sqrt{2\lambda^{2}-1} \over  (2\lambda^{2}-1)^{3}} \,,
\end{eqnarray}
\end{subequations}
recalling that the anisotropy is encoded in $\lambda = \sigma_{1\parallel}/\sigma_{1\perp}$.
The trace of this statistic yields the Minkowski Functional $W_{2}$ in redshift space.

The Gaussian, Kaiser limit of the ensemble averages are 
\begin{subequations}\label{eq:w2g}
\begin{eqnarray}  \langle v_{1}{}^{1} \rangle_\mathrm{G} + \langle v_{2}{}^{2} \rangle_\mathrm{G}  &=& {\sigma_{1\perp}^{2} \over 12\pi\sqrt{2\pi}\sigma^{2}} e^{-\nu^{2}/2} A^{(2)}_{\rm G \perp}H_{1}(\nu)  \,,\\ 
 \langle v_{3}{}^{3} \rangle_\mathrm{G} &=&   {\sigma_{1\perp}^{2} \over 12\pi\sqrt{2\pi}\sigma^{2}} e^{-\nu^{2}/2} A^{(2)}_{\rm G \parallel}H_{1}(\nu) \,.
\end{eqnarray} 
\end{subequations}
We have checked that equation (\ref{eq:w2g}) agrees with previous calculations in redshift space \citep{Appleby_2019}. The trace of (\ref{eq:v11}) should yield the Minkowski Functional $W_{2}$, however the non-Gaussian form of $W_{2}$ was not derived in \citet{Codis:2013exa}. To our knowledge, (\ref{eq:v11}) represents the first derivation of $W_{2}$ to leading non-Gaussian order in redshift space. In the appendix we show that in the isotropic limit, the trace of (\ref{eq:v11}) coincides with $W_{2}$ calculated in \cite{2003ApJ...584....1M}.

To make use of equations~\eqref{eq:w1} and~\eqref{eq:v11} we must express the cosmological dependence of the cumulants $\sigma$, $\sigma_{1\parallel}, \sigma_{1\perp}$, $\lambda$ and $\langle x^{3} \rangle$, $\langle x q_{\perp}^{2}\rangle$, $\langle x x_{3}^{2}\rangle$, $\langle q_{\perp}^{2} J_{1\perp}\rangle $ and $ \langle  x_{3}^{2} J_{1\perp} \rangle $. 
This will be the subject of Section \ref{sec:cumulants}. 
Note importantly that all third order cumulants in equations~\eqref{eq:w1} and~\eqref{eq:v11} are linearly sensitive to the amplitude of matter fluctuations $\sigma$, so measuring the non-Gaussian contribution to the MT curves gives us access to both the growth rate $f\sigma_{8}$ and the amplitude $b_{1}\sigma_{8}$.

\section{Cumulants in Redshift Space}
\label{sec:cumulants}
As the next step, we need to express the two- and three-point cumulants as functions of the cosmological parameters. Fortunately, this has already been done in \cite{Codis:2013exa} for the purpose of calculating the scalar Minkowski Functionals in redshift space; in real space the cumulants up to the trispectrum were constructed in \cite{Matsubara:2020knr}. In this section we review the work of \cite{Codis:2013exa}.

We note that the variables used in this work differ from those in \cite{Appleby_2019,Appleby:2022itn}, where the ensemble expectation values of the MTs in the Gaussian limit were constructed. In those works,  the authors used cumulants that asymptote to their isotropic limits as $f \to 0$. In contrast, here we use precisely the definitions of \cite{Codis:2013exa}, to avoid confusion. We have checked that mapping the conventions in this work yields the same Gaussian limit as \cite{Appleby_2019,Appleby:2022itn}.   
\subsection{Two-point Cumulants} 
The two-point cumulants in redshift space are given by 
\begin{subequations}
\label{eq:ss}
\begin{eqnarray}\label{eq:s0} & & \sigma^{2} = {1 \over (2\pi)^{2}} \int_{-1}^{1}d\mu \int_{0}^{\infty}dk \, k^{2} (1+f\mu^{2})^{2}P_{m}(k,z) W^{2}(kR_{\rm G})\,, \\
\label{eq:s1pa} & & \sigma_{1\parallel}^{2} = {1 \over (2\pi)^{2}} \int_{-1}^{1}d\mu \int_{0}^{\infty}dk \, k^{4} \mu^{2}(1+f\mu^{2})^{2}P_{m}(k,z) W^{2}(kR_{\rm G})\,, \\
\label{eq:s1pe} & & \sigma_{1\perp}^{2} = {1 \over (2\pi)^{2}} \int_{-1}^{1}d\mu \int_{0}^{\infty}dk \, k^{4} (1-\mu^{2})(1+f\mu^{2})^{2}P_{m}(k,z) W^{2}(kR_{\rm G})\,, \\
& & \sigma_{2\parallel}^{2} =   {1 \over (2\pi)^{2}} \int_{-1}^{1}d\mu \int_{0}^{\infty}dk \, k^{6} \mu^{4}(1+f\mu^{2})^{2}P_{m}(k,z) W^{2}(kR_{\rm G})\,, \\
& & \sigma_{2\perp}^{2} = {1 \over (2\pi)^{2}} \int_{-1}^{1}d\mu \int_{0}^{\infty}dk \, k^{6} (1-\mu^{2})^{2}(1+f\mu^{2})^{2}P_{m}(k,z) W^{2}(kR_{\rm G}) \,,\\
& & \sigma_{Q}^{2} = {1 \over (2\pi)^{2}} \int_{-1}^{1}d\mu \int_{0}^{\infty}dk \, k^{6} \mu^{2}(1-\mu^{2})(1+f\mu^{2})^{2}P_{m}(k,z) W^{2}(kR_{\rm G})\,,
\end{eqnarray}
\end{subequations}
where we have introduced the isotropic, linear matter power spectrum $P_{m}(k,z)$ evaluated at redshift $z$, and smoothing kernel $W(k R_{\rm G}) = e^{-k^{2}R_{\rm G}^{2}/2}$ for some comoving smoothing scale $R_{\rm G}$. The $\mu$ integrals can be performed analytically, but in what follows we will introduce an additional $\mu$ dependent kernel to account for the Finger of God effect. We use CAMB\footnote{https://github.com/cmbant/CAMB} to estimate $P_{m}(k,z)$ for a given cosmology and define the $i^{\rm th}$ isotropic cumulant as 
\begin{equation} \sigma_{i}^{2} = {1 \over 2\pi^{2}} \int_{0}^{\infty} dk k^{2i+2} P_{m}(k, z)  W^{2}(kR_{\rm G})  . \end{equation}

\subsection{Three-point Cumulants}
The statistic $\langle w_{i}{}^{j} \rangle$ is sensitive to the cubic cumulants $\langle x^{3}\rangle$, $\langle x x_{I}^{2} \rangle$ and $\langle x x_{3}^{2}\rangle$. The second MT $\langle v_{i}{}^{j} \rangle$ is also sensitive to $\langle q_{\perp}^{2} J_{1\perp}\rangle$ and $\langle x_{3}^{2} J_{1\perp}\rangle$. By utilising isotropy in the plane perpendicular to the line of sight, the cumulants $\langle x x_{I} x_{3}\rangle$ and $\langle x x_{I} x_{J} \rangle$, $I \neq J$ are zero. 

For a dark matter field, the non-zero cumulants are given by \cite{Codis:2013exa}; they generically obey
\begin{equation}
\label{eq:x3}\hskip -0.55cm \langle {\cal Y} \rangle \!=\!\! \int\!\! {d^{3}k_{1} \over (2\pi)^{3}}\!{d^{3}k_{2} \over (2\pi)^{3}}\!\left(\! {b_{2} \over 2} Z^{2}_{1}({\bf k}_{1}) P_{m}(k_{1},z)P_{m}(k_{2},z)\alpha({\bf k}_{1},{\bf k}_{2})\! +\! 2 Z_{1}({\bf k}_{1}) Z_{1}({\bf k}_{2}) Z_{2}({\bf k}_{1},{\bf k}_{2}) P_{m}(k_{1},z)P_{m}(k_{2},z)\beta({\bf k}_{1},{\bf k}_{2})\!  \right)\!, 
\end{equation} 
 where ${\cal Y} = x^{3}, xq_{\perp}^{2}, x x_{3}^{2}, q_{\perp}^{2}J_{1\perp}$, and $x_{3}^{2}J_{1\perp}$ resp. 
 The functions $\beta({\bf k}_{1},{\bf k}_{2})$ for each ${\cal Y}$ are given in Table \ref{tab:1},
\begin{table}[ht]
\begin{center}
\begin{tabular}{ |c|c|c| }
 \hline
 ${\cal Y}$ & $\alpha({\bf k}_{1},{\bf k}_{2})$ &  $\beta({\bf k}_{1},{\bf k}_{2})$ 
\\ \hline 
 $x^{3}$ & $3/\sigma^{3}$  & $3/\sigma^{3}$ \\ 
 $x q_{\perp}^{2}$ & $k_{1\perp}^{2}/(\sigma \sigma_{1\perp}^{2})$  & $[{\bf k}_{1\perp} . {\bf k}_{2\perp} + 2 k_{1\perp}^{2}]/(\sigma \sigma_{1\perp}^{2})$ \\ 
 $x x_{3}^{2}$ & $k_{1\parallel}^{2}/(\sigma \sigma_{1\parallel}^{2})$  & $[k_{1\parallel}k_{2\parallel} + 2k_{1\parallel}^{2}]/(\sigma \sigma_{1\parallel}^{2})$ \\
 $q_{\perp}^{2}J_{1\perp}$ & 0 & $[({\bf k}_{1\perp} . {\bf k}_{2\perp})^{2} - ({\bf k}_{1\perp} \times {\bf k}_{2\perp})^{2} - k_{1\perp}^{2}k_{2\perp}^{2}]/(\sigma_{1\perp}^{2}\sigma_{2\perp})$
  \\ 
$x_{3}^{2} J_{1\perp}$   & 0  &  $[ k_{1\parallel}k_{2\parallel}({\bf k}_{1\perp} + {\bf k}_{2\perp})^{2} - 2k_{1\perp}^{2}k_{2\parallel}(k_{1\parallel} + k_{2\parallel})]/(\sigma_{1\parallel}^{2}\sigma_{2\perp})$ \\ 
 \hline
\end{tabular}
\end{center}
\caption{\label{tab:1}Bispectrum kernels for each of the cumulants pertinent to $\langle w_{i}{}^{j}\rangle$ and $\langle v_{i}{}^{j}\rangle$.  }
\end{table}
%
while the redshift space kernels $Z_1$ and $Z_2$ are
\begin{subequations}
\begin{eqnarray} & & Z_{1}({\bf k}_{i}) = b_{1} + f \mu^{2}_{i}\,, \\
& & Z_{2}({\bf k}_{i}, {\bf k}_{j}) = b_{1}F_{2}({\bf k}_{i}, {\bf k}_{j}) + {b_{2} \over 2} + f \mu_{ij}^{2} G_{2}({\bf k}_{i}, {\bf k}_{j}) + {f \mu_{ij} k_{ij} \over 2} b_{1} \left({\mu_{i} \over k_{i}} + {\mu_{j} \over k_{j}} \right) + {(f\mu_{ij} k_{ij})^{2} \over 2} {\mu_{i} \over k_{i}}{\mu_{j} \over k_{j}} \,, \\
& & F_{2}({\bf k}_{i}, {\bf k}_{j}) = {5 \over 7} + {m_{ij} \over 2}\left({k_{i} \over k_{j}} + {k_{j} \over k_{i}}\right) + {2 \over 7} m_{ij}^{2}\,, \\ 
& & G_{2}({\bf k}_{i}, {\bf k}_{j}) = {3 \over 7} + {m_{ij} \over 2} \left({k_{i} \over k_{j}} + {k_{j} \over k_{i}}\right) + {4 \over 7} m_{ij}^{2} \,,
\end{eqnarray} 
\end{subequations}
 where $m_{ij} = {\bf k}_{i} \cdot  {\bf k}_{j} /(k_{i}k_{j})$, ${\bf k}_{ij} = {\bf k}_{i} + {\bf k}_{j}$, $\mu_{ij} = {\bf k}_{ij} \cdot {\bf e}_{3}/k_{ij} = (\mu_{i}k_{i} + \mu_{j}k_{j})/k_{ij}$, $\mu_{i} = {\bf k}_{i}\cdot{\bf e}_{3}/k_{i}$. Although we focus on dark matter in this work, we include $b_{1}$ and $\,b_{2}$ in the cumulants, which are respectively the first and second order bias parameters and will be pertinent to galaxy data in the future. In what follows we fix $b_{1} = 1$ and $b_{2} = 0$.

To proceed with integration in equation~(\ref{eq:x3}) in redshift space 
we decompose the Fourier modes into vectors projected perpendicular and parallel to the line of sight: ${\bf k} = {\bf k}_{\parallel} + {\bf k}_{\perp}$ with ${\bf k}_{\perp} \cdot {\bf e}_{3} = 0$, ${\bf k}_{\parallel} \cdot {\bf e}_{3} = k_{\parallel}$, ${\bf k}_{\parallel} \cdot {\bf k}_{\perp} = 0$ and $\theta_{12}$, the angle between ${\bf k}_{1\perp}$ and ${\bf k}_{2\perp}$ in the plane perpendicular to ${\bf e}_{3}$. In terms of magnitudes and angles defined above, we can write $k_{i,\parallel} = \mu_{i}k_{i}$, $k_{i, \perp} = \sqrt{1-\mu_{i}^{2}}k_{i}$, and ${\bf k}_{i, \perp}  \cdot {\bf k}_{j, \perp} = \sqrt{1-\mu_{i}^{2}}\sqrt{1-\mu_{j}^{2}}k_{i}k_{j}\cos\theta_{ij}$ and  
\begin{subequations}
    \begin{eqnarray} & & k_{i} = \sqrt{k_{i, \parallel}^{2} + k_{i, \perp}^{2}}\,, 
\quad \mu_{i} = k_{i, \parallel}/k_{i} \,, \\
& & k_{ij} = \sqrt{k_{i,\parallel}^{2} + k_{i,\perp}^{2} + k_{j,\parallel}^{2} + k_{j,\perp}^{2} + 2 k_{i,\parallel}k_{j,\parallel} + 2{\bf k}_{i,\perp}\cdot{\bf k}_{j,\perp}}\,,\\
& & \mu_{ij} = (k_{i,\parallel} + k_{j,\parallel})/k_{ij}\,, 
\quad m_{ij}=(k_{i,\parallel}k_{j,\parallel}+{\bf k}_{i,\perp} \cdot {\bf k}_{j,\perp})/(k_{i}k_{j})\,.
\end{eqnarray} 
\end{subequations}
In redshift space, five of the six Fourier mode integrations are non-trivial. There is a single trivial integration, which is one of the angles in the plane perpendicular to the line of sight. Specifically, only the relative angle between ${\bf k}_{i,\perp}$ and ${\bf k}_{j,\perp}$ enters into the bispectrum kernels, because the field is isotropic in this plane.  Thus, with the inclusion of the smoothing kernels, we can present the Fourier integrals as follows
\begin{eqnarray} \nonumber \int {d^{3}k_{1} \over (2\pi)^{3}}{d^{3}k_{2} \over (2\pi)^{3}} &\to& {1 \over (2\pi)^{5}}\int k_{1}^{2} k_{2}^{2}dk_{1}dk_{2} d\mu_{1}d\mu_{2} d\theta_{12} W_{G}\left(k_{1}R_\mathrm{G}\right)W_{G}\left(k_{2}R_\mathrm{G}\right) W_{G}\left({|{\bf k}_{1}+{\bf k}_{2}|R_\mathrm{G}}\right)\,,
\end{eqnarray} 

\subsection{Finger of God Effect}

Beyond the perturbative non-Gaussianity generated by gravitational collapse, there is an additional effect due to stochastic velocities of bound systems scattering the redshift-space positions of tracer particles along the line of sight. This Finger-of-God velocity dispersion can be significant even on relatively large scales \citep{Juszkiewicz:1998em,Hikage:2013yja,10.1093/mnras/stu1051,10.1093/mnras/stu1391,Tonegawa:2020wuh,Okumura:2015fga}; this was also observed in \cite{Appleby:2022itn}. To account for the effect, we introduce an additional $\mu$-dependent function in the two- and three- point cumulant definitions  \citep{1976Ap&SS..45....3P,Peacock:1993xg,1994ApJ...431..569P,Desjacques:2009kt,PhysRevD.70.083007}. Specifically, the smoothing kernel in the two-point cumulants (\ref{eq:ss}) is modified to
\begin{equation}
 W^{2}(kR_{\rm G}) \to  W^{2}(kR_{\rm G}) e^{-k^{2} \mu^{2} \sigma_{v}^{2}}\,,
\end{equation}
and the three-point cumulant integrand (\ref{eq:x3}) is modified to (remembering that we are fixing $b_{2} = 0$)  
\begin{equation}
\label{eq:bis}  2 \left[ Z_{2}({\bf k}_{1},{\bf k}_{2}) Z_{1}({\bf k}_{1})Z_{1}({\bf k}_{2}) P_{m}(k_{1},z)P_{m}(k_{2},z)e^{-(k_{1}^{2}\mu_{1}^{2} + k_{2}^{2} \mu_{2}^{2}) \sigma_{B}^{2}}  \right]  \,, 
\end{equation}
 with two free parameters $\sigma_{v}$ and $\sigma_{B}$. This is a phenomenological parameterization, and as we will see the presence of stochastic velocity dispersion disproportionately affects the components of $\langle w_{i}{}^{j} \rangle$, $\langle v_{i}{}^{j} \rangle$ parallel to the line of sight. Due to the phenomenological nature of the prescription, the parameters $\sigma_{v}$, $\sigma_{B}$ are sometimes treated as independent parameters to be jointly constrained. In this work we always fix them to be equal.
 \subsection{Shot Noise}
 We can add shot noise to the two-point cumulants by modifying the matter power spectrum according to $P_{m}(k,z) \to P_{m}(k,z) + \bar{n}^{-1}$, where $\bar{n}$ is the number density of the point distribution. Similarly, we can add shot noise to the bispectrum. Specifically the integrand defined in (\ref{eq:bis}) picks up an additive correction of the form \citep{10.1093/mnras/290.4.651}
\begin{equation}
B_{\rm shot}(k_{1},k_{2},|{\bf k}_{1}+{\bf k}_{2}|,z) = \epsilon_{0} + \eta_{0}\left[ P_{m}(k_{1},z) + P_{m}(k_{2},z) + P_{m}(|{\bf k}_{1}+{\bf k}_{2}|,z) \right] \,.
\end{equation} 
If we assume that the shot noise is Poissonian, then $\epsilon_{0} = 1/\bar{n}^{2}$ and $\eta_{0} = 1/\bar{n}$, and $P_{m}(k,z)$ is the power spectrum without any shot noise included. However, in this work we focus on dark matter and smooth on scales much larger than the mean particle separation, meaning that shot noise will be completely suppressed. Practically, when $\bar{n}^{-1/3} \ll R_{\rm G}$ these contributions can be neglected. 
\section{Data and Methodology}

To confirm the results derived in the previous section, we compare $\langle w_{i}{}^{j}\rangle$ and $\langle v_{i}{}^{j} \rangle$ to numerical reconstructions of $W^{0,2}_{1}$ and $W^{0,2}_{2}$. In this section we review the mock data and numerical algorithms used to estimate $W^{0,2}_{1}$ and $W^{0,2}_{2}$. We direct the reader to \cite{1367-2630-15-8-083028,JMI:JMI3331,Collischon:2024jhw} for related numerical studies of the Minkowski tensors.

To study redshift-space Minkowski tensors of the non-Gaussian dark matter density field in simulations, we use $N_{\rm real} = 100$ number of realisations of $z=1$ snapshot boxes from the Quijote simulations \citep{Villaescusa-Navarro:2019bje}). These are a suite of dark matter simulations in which $\sim 44,000$ realisations of $512^{3}$ particles are gravitationally evolved in boxes of size $L = 1000 \, h^{-1} {\rm Mpc}$ from $z=127$ to $z=0$. We use data from the fiducial cosmology, with parameters $[\Omega_{m}, \Omega_{b}, h, n_{s}, \sigma_{8}] = [0.3175, 0.049, 0.6711, 0.9624, 0.834]$. 

To generate the plane-parallel redshift space distorted fields, we define the redshift space positions of the dark matter particles $({\bf s})$ by perturbing their real-space positions $({\bf x})$ according to 
\begin{equation}  
{\bf s} = {\bf x} +  {\bf e}_{3} ({\bf v}\cdot{\bf e}_{3}) {(1+z) \over H(z)} , 
\end{equation} 
where ${\bf v}$ is the velocity of the particle, ${\bf e}_{3}$ is the unit vector aligned with the $x_{3}$ direction of the Cartesian grid that is taken to be the line of sight and we take $z=1$. Periodicity is enforced after the redshift space correction. There are two reasons why we make our measurements at $z=1$. First, perturbation theory is applicable on relatively smaller scales at higher redshift. Second, modern data sets typically observe galaxies at $z \sim 1$, making this regime more pertinent than $z=0$ data. 

After moving the dark matter particles to their redshift space position, we bin them into pixels using a cloud-in-cell scheme, using a regular Cartesian lattice with $N=256$ pixels per side of resolution $\Delta = 1000/256 = 3.9 \, h^{-1} \, {\rm Mpc}$ and define the number density field $\delta_{ijk} = (n_{ijk} - \bar{n})/\bar{n}$. We smooth these fields with Gaussian kernel $W(kR_{\rm G}) = e^{-k^{2}R_\mathrm{G}^{2}/2}$ in Fourier space, calculate the mean $\bar{\delta}$ and variance $\sigma^{2}$ after smoothing, and finally construct the zero mean, unit variance density field $\delta_{ijk} \to (\delta_{ijk} - \bar{\delta})/\sigma$. The quantities $W^{0,2}_{1}|_{i}{}^{j}$ and $W^{0,2}_{2}|_{i}{}^{j}$ for this field are then measured over $n_{\nu} = 101$ values of threshold density $-3 \leq \nu \leq 4.5$. We repeat the measurements for smoothing scales $R_{\rm G} = 15, 20, 25, 30 \, h^{-1} \, {\rm Mpc}$. 

To numerically extract the Minkowski Tensors from the density fields, we use the method developed in \cite{2018ApJ...863..200A}. Briefly, we construct an iso-field triangulated surface mesh by generating a tetrahedral mesh from the pixels, and use the triangulated mesh and their unit normal vectors to construct the tensors \citep{1367-2630-15-8-083028}  
\begin{eqnarray}
 \label{eq:num1} & & W^{0,2}_{1}|_{i}{}^{j} = {1 \over 6V} \sum_{t} A_t{\bf \hat{n}}_{i}{\bf \hat{n}}^{j} \,,\\
 \label{eq:num2} & & W^{0,2}_{2}|_{i}{}^{j} = {1 \over 3\pi V} \sum_{k} |{\bf e}_{k}|\left( (\alpha_{k} + \sin \alpha_{k})({\bf \ddot{n}}_{k}^{2})_{i}{}^{j} + (\alpha_{k}-\sin \alpha_{k}) ({\bf \dot{n}}_{k}^{2})_{i}{}^{j}\right)\,,
\end{eqnarray}
where the sum in equation~\eqref{eq:num1} is over all triangles in the surface mesh and ${\bf \hat{n}}_{i}$ the corresponding unit normal vectors pointing externally to the surface, and $A_{t}$ is the area of the $t^{\rm th}$ triangle. In equation~\eqref{eq:num2}, the sum is over $k$ unique triangle edges in the mesh and $|{\bf e}_{k}|$ is the length of the $k^{\rm th}$ edge. Each edge is shared by two triangles with unit normals ${\bf \hat{n}}_{k}$ and ${\bf \hat{n}}'_{k}$, $\alpha_{e}$ is the angle sub-tending these normals and we define ${\bf \ddot{n}}_{k} = ({\bf \hat{n}}_{k} + {\bf \hat{n}}'_{k})/|{\bf \hat{n}}_{k} + {\bf \hat{n}}'_{k}|$, ${\bf \dot{n}}_{k} = {\bf \ddot{n}}_{k} \times {\bf e}_{k}$. The square of a vector in the above expression is shorthand for the symmetric tensor product and $V = (1000 \, h^{-1} \, {\rm Mpc})^{3}$ is the total volume occupied by the field.

\section{Numerical Reconstruction of Cumulants and Minkowski Tensors}
\label{sec:results}

\begin{figure}
    \centering
    \includegraphics[width=0.98\textwidth]{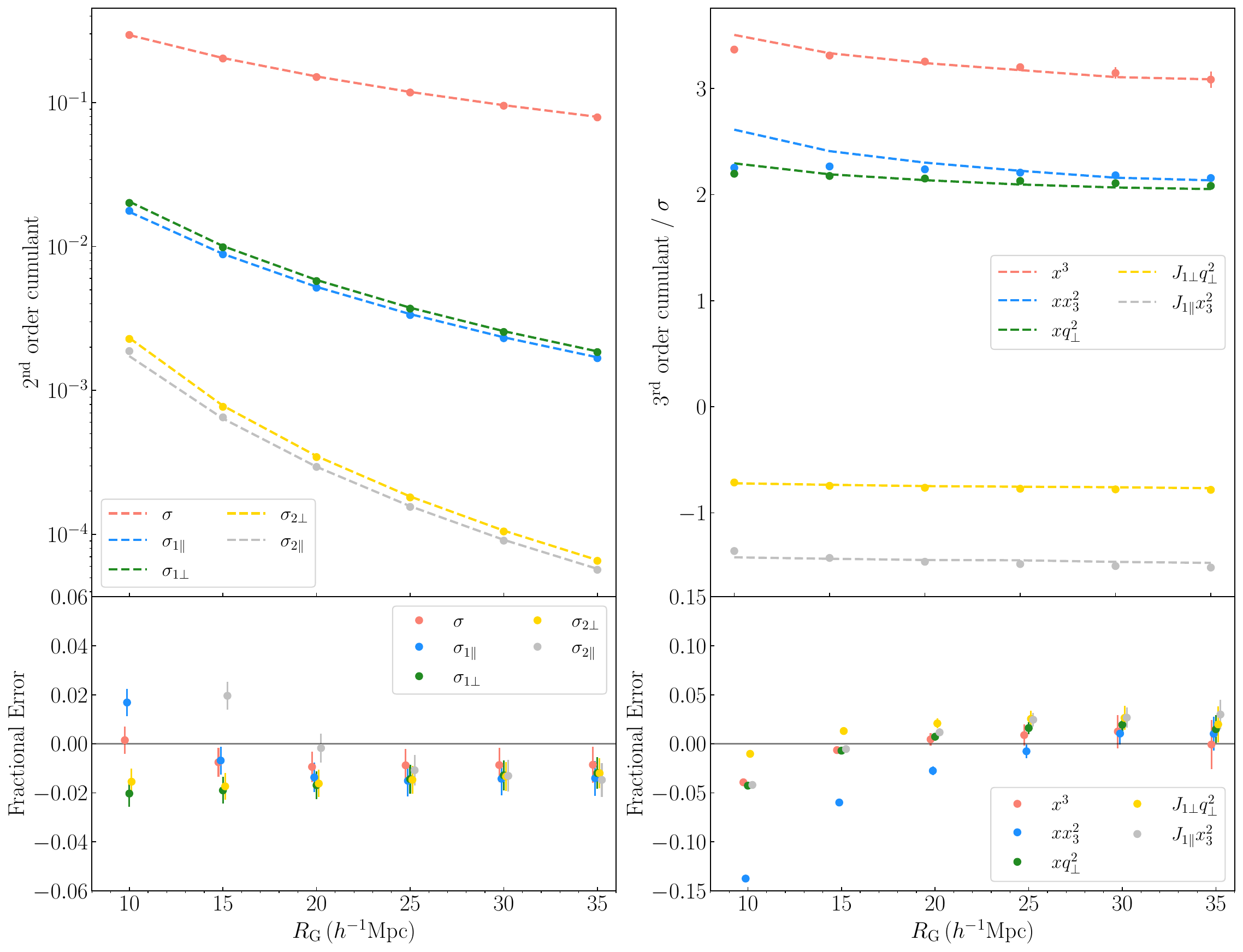}
    \caption{[Top panels] Numerically measured two-point (left panel) and three-point (right panel) cumulants from a set of $z=1$ dark matter snapshot boxes in redshift space (points/error bars). The dashed lines are the corresponding ensemble expectation values constructed in Section \ref{sec:cumulants}, taking $\sigma_{v} = \sigma_{B} = 4.9 \, h^{-1} \, {\rm Mpc}$. Error bars are the error on the mean. The scaling of most cumulants is well captured by perturbation theory, notwithstanding  a small
    departure at smaller scales for $x x_3^2$. [Bottom panels] Fractional uncertainty between the measured and theoretical estimates of the two-point (left panel) and three-point (right panel) cumulants. The agreement is percent level for $R_{G} > 20 \, h^{-1} \, {\rm Mpc}$.  }
    \label{fig:0}
\end{figure}

First, we measure the two-point ($\sigma$, $\sigma_{1\perp}$, $\sigma_{1\parallel}$, $\sigma_{2\perp}$, $\sigma_{2\parallel}$) and three point $(\langle x^{3} \rangle$, $\langle x q_{\perp}^{2} \rangle$, $\langle x x_{3}^{2} \rangle$, $\langle x_{3}^{2} J_{1\perp} \rangle$, $\langle q_{\perp}^{2} J_{1\perp}\rangle)$ cumulants from the simulations, and compare them to the expected values computed according to the perturbative formalism of Section \ref{sec:cumulants}. The results are presented in Figure \ref{fig:0}, as a function of smoothing scale $R_\mathrm{G}$. The top left/right panels are the two/three point cumulants, and the points/error bars are mean and error on mean measured from the snapshot boxes, while the dashed lines are obtained using perturbation theory. To calculate the cumulants using the results of Section \ref{sec:cumulants} we adopt the input fiducial cosmological parameters and also take FoG velocity dispersion $\sigma_{v} = \sigma_{B} = 4.9 \, h^{-1} \, {\rm Mpc}$. We explain how this value was determined in Appendix \ref{sec:app_b}. The agreement is excellent for all cumulants for scales $R_\mathrm{G} \geq 20 \, h^{-1} \, {\rm Mpc}$, with the three point cumulants starting to deviate for $R_{\rm G} \sim 15 \, h^{-1} \, {\rm Mpc}$. In the bottom panels of the Figure, we present the fractional difference between the measured and theoretical cumulants, divided by the theoretical prediction. The points/error bars are again the mean and error on the mean. We find percent level agreement between theory and measurement for $R_{G} > 20 \, h^{-1} \, {\rm Mpc}$, with a small (subpercent) systematic difference in the two point cumulants (cf bottom left panel) on large scales. We note that the three point cumulants, divided by $\sigma$, are only very weakly sensitive to the smoothing scale $R_\mathrm{G}$, a result that is consistent with \cite{Codis:2013exa,2003ApJ...584....1M}. 

\begin{figure}
    \centering
    \includegraphics[width=0.98\textwidth]{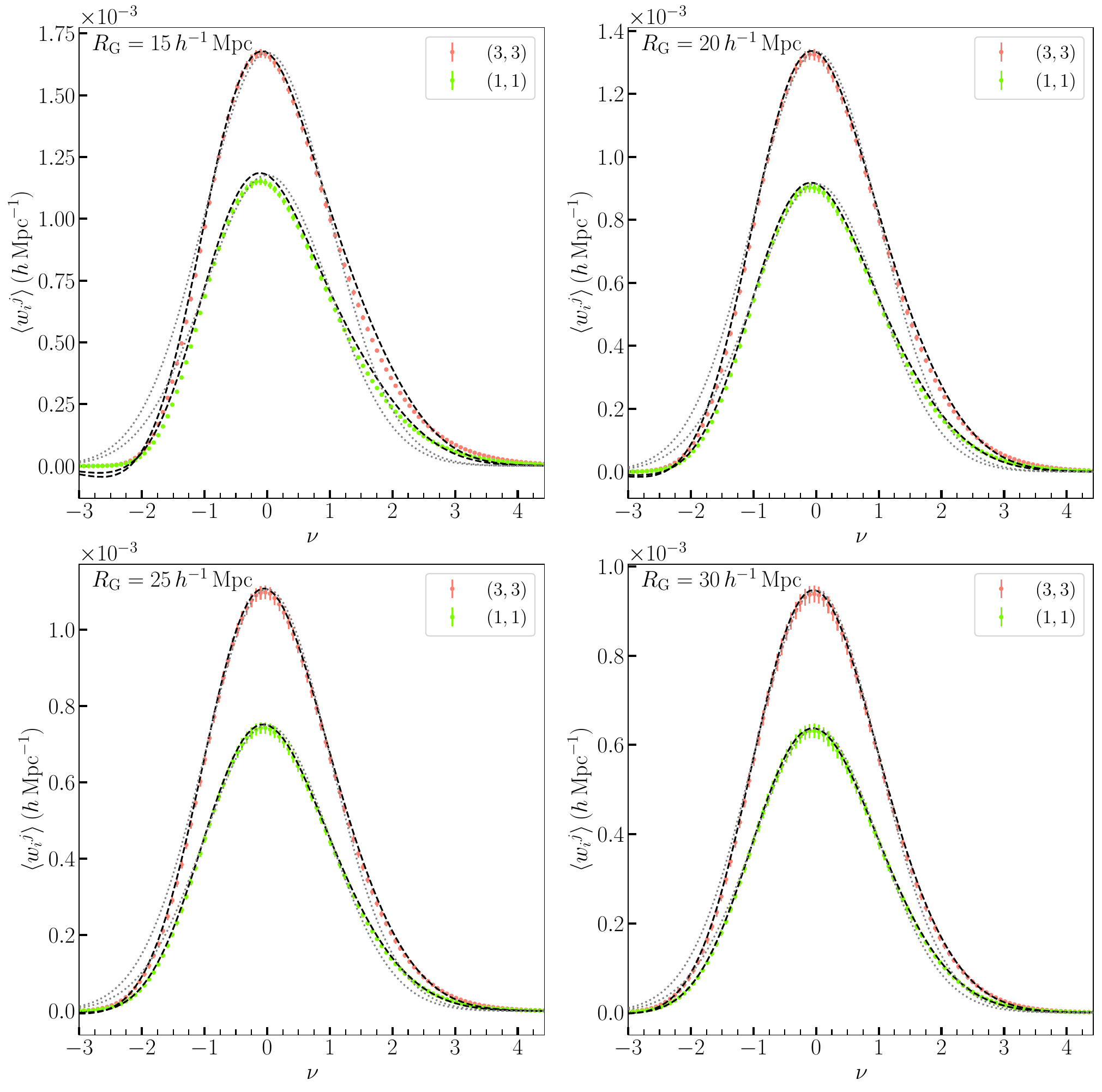}
    \caption{The Minkowski tensor components $W_{1}^{0,2}|_{1}{}^{1}$ and $W_{1}^{0,2}|_{3}{}^{3}$ (green/red points and error bars) extracted from $N=100$ realisations of quijote $z=1$ dark matter snapshot boxes in redshift space. The black dashed lines are the theoretical predictions for the ensemble averages $\langle w_{1}{}^{1}\rangle$ and $\langle w_{3}{}^{3}\rangle$ inferred from  Section \ref{sec:ens_w}, with Finger of God velocity dispersion $\sigma_{v} = \sigma_{B} = 4.9 \, h^{-1} \, {\rm Mpc}$. The grey dotted lines are the Gaussian predictions $\langle w_{1}{}^{1}\rangle_\mathrm{G}$ and $\langle w_{3}{}^{3}\rangle_\mathrm{G}$ with the same FoG velocity dispersion included.  The top left/right panels correspond to smoothing scales $R_{\rm G} = 15, 20 \, h^{-1} \, {\rm Mpc}$ and the bottom left/right panels $R_{\rm G} = 25, 30 \, h^{-1} \, {\rm Mpc}$. On the range of considered scales, the non gaussian features are significant and well captured by perturbation theory.
    See also Fig~\ref{fig:app1} without the FoG correction.}
    \label{fig:1b}
\end{figure}

In Figure \ref{fig:1b} we present measurements of the Minkowski tensor components $W^{0,2}_{1}|_{1}{}^{1}$, $ W^{0,2}_{1}|_{3}{}^{3}$ (green, red points/error bars) extracted from the $z=1$ redshift space snapshot boxes. The $W^{0,2}_{1}|_{2}{}^{2}$ component is statistically indistinguishable from $W^{0,2}_{1}|_{1}{}^{1}$ and not plotted, and all off-diagonal elements are consistent with zero and also not plotted. Each panel corresponds to a different smoothing scale $R_\mathrm{G} = 15, 20, 25, 30 \, h^{-1} \, {\rm Mpc}$. The black dashed lines are the corresponding theoretical expectation values $\langle w_{1}{}^{1} + w_{2}{}^{2} \rangle/2$ and $\langle w_{3}{}^{3} \rangle$ from equations (\ref{eq:w1}) using the two- and three-point cumulants inferred from perturbation theory. The faint grey dotted lines are the Gaussian limits of the perturbation theory predictions $\langle w_{1}{}^{1} + w_{2}{}^{2} \rangle_\mathrm{G}/2$ and $\langle w_{3}{}^{3} \rangle_\mathrm{G}$, obtained by setting all three-point cumulants to zero. The agreement between perturbation theory and measurements (black dashed lines and green/red points) is good for scales $R_\mathrm{G} > 20 \, h^{-1} \, {\rm Mpc}$, but higher order contributions increasingly modify the shape of the curves on smaller scales (cf top left panel). For large negative thresholds, the ensemble average predicts that $\langle w_{i}{}^{i} \rangle$ is negative, despite these quantities being positive definite by definition. This indicates a failure of both perturbation theory and the Edgeworth expansion in this regime, to the order at which we are working. The small scale breakdown of perturbation theory is clearly presented in the lower panels of Figure \ref{fig:0}, where the theoretical predictions depart from the measured values of the cumulants for $R_{G} < 20 \, h^{-1} \, {\rm Mpc}$. This is particularly apparent for the bispectrum components $x^{3}$, $x x_{3}^{2}$ and $x q_{\perp}^{2}$ (cf. bottom right panel). Independently, the breakdown of the Edgeworth expansion is a known phenomenon \citep{Sellentin:2017aii}, and can lead to spurious negative probability density functions and summary statistics, as we see in the top panels of Figure \ref{fig:1b}. The breakdown of the Edgeworth expansion can be studied using fields for which non-perturbatively non-Gaussian analytic results are known \citep{Bernardeau:2024ysi,Bernardo:2025srg}.

\begin{figure}
    \centering
    \includegraphics[width=0.98\textwidth]{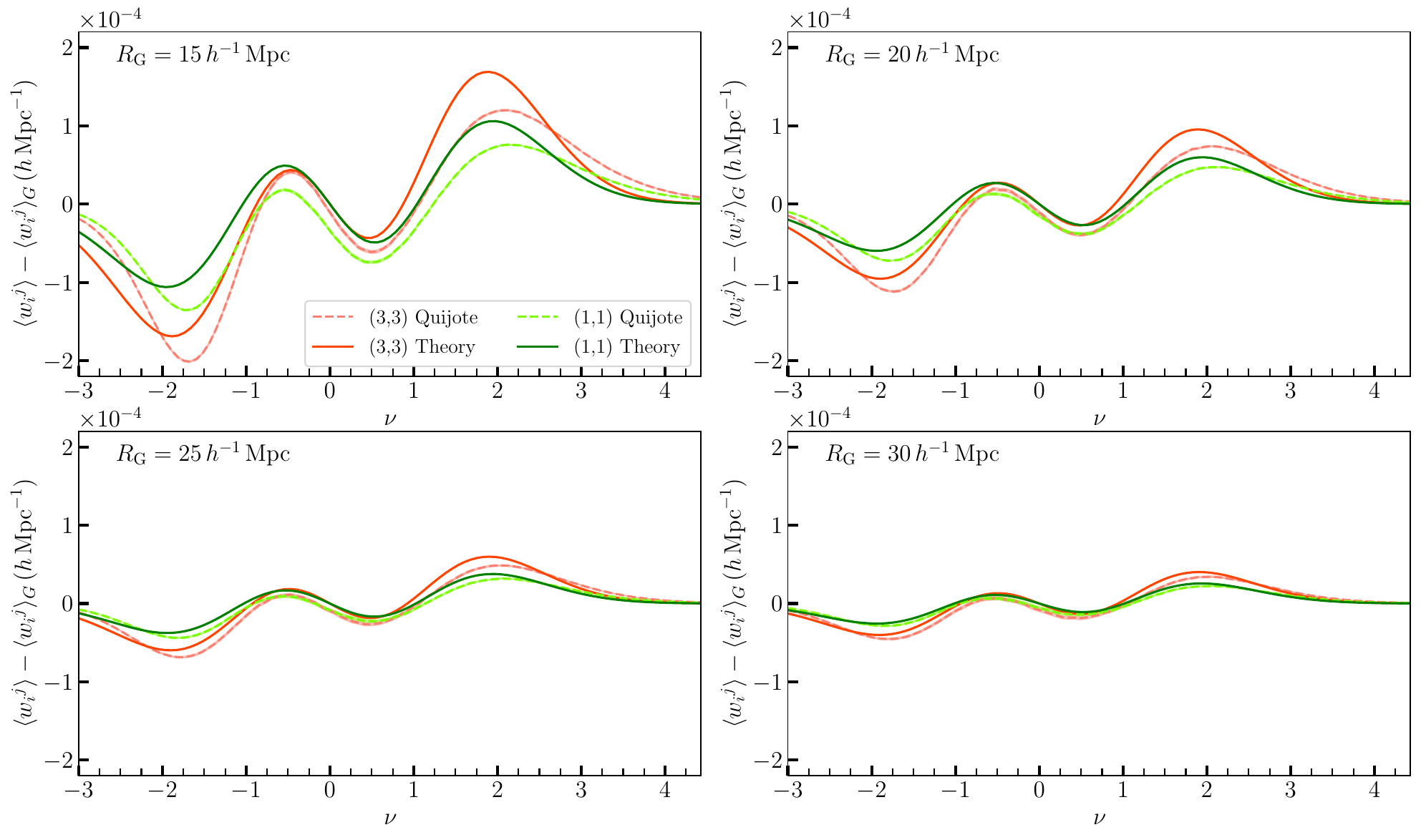}
    \caption{Residuals  $W_{1}^{0,2}|_{i}{}^{i} - \langle w_{i}{}^{i} \rangle_\mathrm{G}$ (red/green dashed lines) and $\langle w_{i}{}^{i}\rangle - \langle w_{i}{}^{i} \rangle_\mathrm{G}$ (red/green solid lines), where $W_{1}^{0,2}|_{i}{}^{i}$ is the mean value of the Minkowski tensor components from the snapshot boxes, $\langle w_{i}{}^{i}\rangle$ is the non-Gaussian prediction obtained in this work and $\langle w_{i}{}^{i} \rangle_\mathrm{G}$ the Gaussian limit of the prediction. The residuals are dominated by the odd Hermite polynomials $H_{1}(\nu)$ and $H_{3}(\nu)$.  As expected the larger the smoothing scale, the better the match and the smaller the amplitude.
    }
    \label{fig:res1}
\end{figure}

In Figure \ref{fig:res1} we present the difference between the measured values of the Minkowski tensor components and the Gaussian expectation values (green/red dashed lines). We present the error on the mean as a shaded region around the dashed lines; its width is comparable to the thickness of the lines. The solid green/red lines are the residuals between the predictions (equation~\ref{eq:w1}) and their Gaussian limit with all cubic cumulants zero. We observe a characteristic pattern in the residuals, which predominantly arises due to the odd Hermite polynomials $H_{1}(\nu)$ and $H_{3}(\nu)$. The agreement between the prediction and measurements (solid/dashed lines) improves as the smoothing scale increases. However, we do not expect these curves to agree exactly, as the measured components (dashed lines) also contain all higher point information whereas the analytic calculation is truncated at the three-point level. The difference between the solid and dashed curves is predominantly due to the four-point cumulants, and should be accounted for when performing cosmological parameter estimation using the MTs.  Ideally one would like to express the quartic cumulants in terms of the trispectrum using perturbation theory, in which case the four point contributions could also be leveraged to constrain cosmological parameters. However, in lieu of an analytic prediction for the higher order cumulants, one approach would be to include additional Hermite coefficients as free parameters and marginalise over them. For example one could fit the following function to measured $W_{1}^{0,2}$ curves,
\begin{subequations}
\begin{align} \nonumber \langle w_{1}{}^{1} \rangle + \langle w_{2}{}^{2} \rangle  &= {\sigma_{1\perp} e^{-\nu^{2} /2}\over 6\pi \sigma  }\biggl[ A^{(1)}_{\rm G \perp}\left(H_{0}(\nu) + {\langle x^{3} \rangle \over 6} H_{3}(\nu)\right) + \bigg( B^{(1)}_{\perp}\langle x q_{\perp}^{2} \rangle  + C^{(1)}_{\perp} \langle x x_{3}^{2} \rangle   \bigg) H_{1}(\nu)    \\ 
&\phantom{= \frac{\sigma_{1\perp} e^{-\nu^{2}/2}}{6\pi\sigma} \bigl[} + h_{6\perp}H_{6}(\nu) + h_{4\perp}H_{4}(\nu) + h_{2\perp}H_{2}(\nu)  \biggr]  \,, \\
\nonumber  \langle w_{3}{}^{3} \rangle &=   {\sigma_{1\perp} e^{-\nu^{2} /2}\over 6\pi \sigma  }\biggl[ A^{(1)}_{\rm G \parallel}\left(H_{0}(\nu) + {\langle x^{3} \rangle \over 6} H_{3}(\nu)\right) + \bigg( B^{(1)}_{\parallel}\langle x q_{\perp}^{2} \rangle  + C^{(1)}_{\parallel} \langle x x_{3}^{2} \rangle   \bigg) H_{1}(\nu)  \\
&\phantom{= \frac{\sigma_{1\perp} e^{-\nu^{2}/2}}{6\pi\sigma} \bigl[} + h_{6\parallel}H_{6}(\nu) + h_{4\parallel}H_{4}(\nu) + h_{2\parallel}H_{2}(\nu)  \biggr] \,,
\end{align}
\end{subequations}
where $h_{2,4,6\perp}$ and $h_{2,4,6\parallel}$ are free parameters to be marginalised over\footnote{This approach is complicated by the fact that the quartic cumulants will also contribute a correction to the coefficients of $H_{0}(\nu)$; the amplitude of the MT curves. Additional terms $h_{0\perp}H_{0}(\nu)$ and $h_{0\parallel}H_{0}(\nu)$ could also be included, although the prior ranges of $h_{0\perp}, h_{0\parallel}$ should be carefully restricted.}. This would account for the contribution of the fourth order cumulants to the shape of $W_{1}^{0,2}$, and higher order contributions could be similarly accounted for at a cost of increasing the number of nuisance parameters $h_{i\perp}$, $h_{i\parallel}$. The underlying point is that setting all higher order cumulants to zero in the ensemble average will lead to a worse fit and potentially parameter biases. A more detailed exploration of this issue will be pursued in future work. 

\begin{figure}
    \centering
    \includegraphics[width=0.98\textwidth]{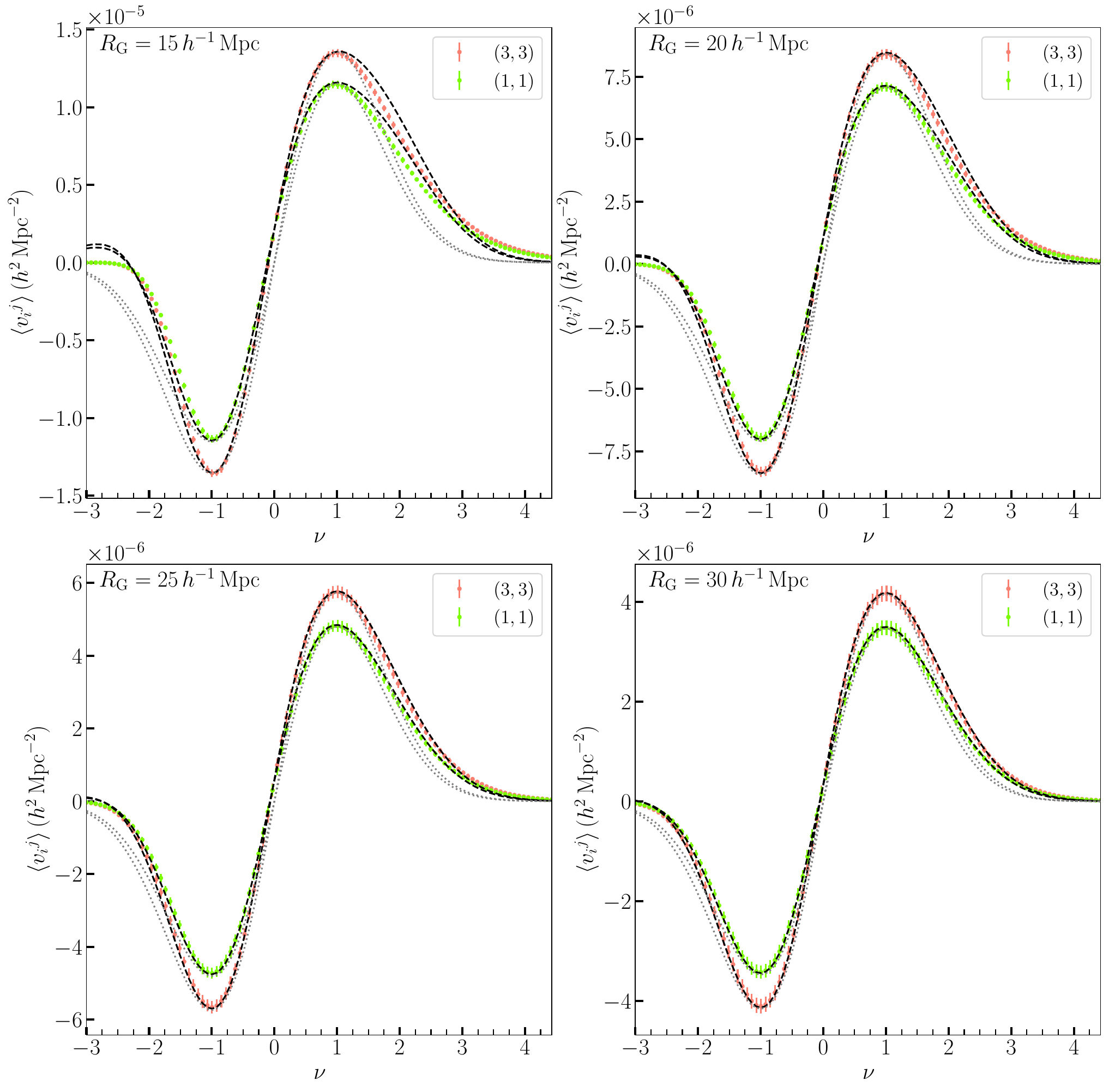}
    \caption{The Minkowski tensor components $W_{2}^{0,2}|_{1}{}^{1}$ and $W_{2}^{0,2}|_{3}{}^{3}$ (green/red points and 1$\sigma$ error bars) extracted from $N=100$ realisations of quijote $z=1$ dark matter snapshot boxes in redshift space. The black dashed lines are the theoretical predictions for the ensemble averages $\langle v_{1}{}^{1}\rangle$ and $\langle v_{3}{}^{3}\rangle$ inferred from  Section \ref{sec:ens_v}, with Finger of God velocity dispersion $\sigma_{v} = \sigma_{B} = 4.9 \, h^{-1} \, {\rm Mpc}$. The grey dotted lines are the Gaussian predictions with the same FoG velocity dispersion included.  The top left/right panels correspond to smoothing scales $R_{\rm G} = 15, 20 \, h^{-1} \, {\rm Mpc}$ and the bottom left/right panels $R_{\rm G} = 25, 30 \, h^{-1} \, {\rm Mpc}$. 
    As for $W_{1}^{0,2}|_{i}{}^{i}$ (Figure~\ref{fig:1b}), the components $W_{2}^{0,2}|_{1}{}^{1}$
    display informative non Gaussian signature,  which are best captured by 
    perturbation theory on larger scales. 
    }
    \label{fig:1c}
\end{figure}

\begin{figure}
    \centering
    \includegraphics[width=0.98\textwidth]{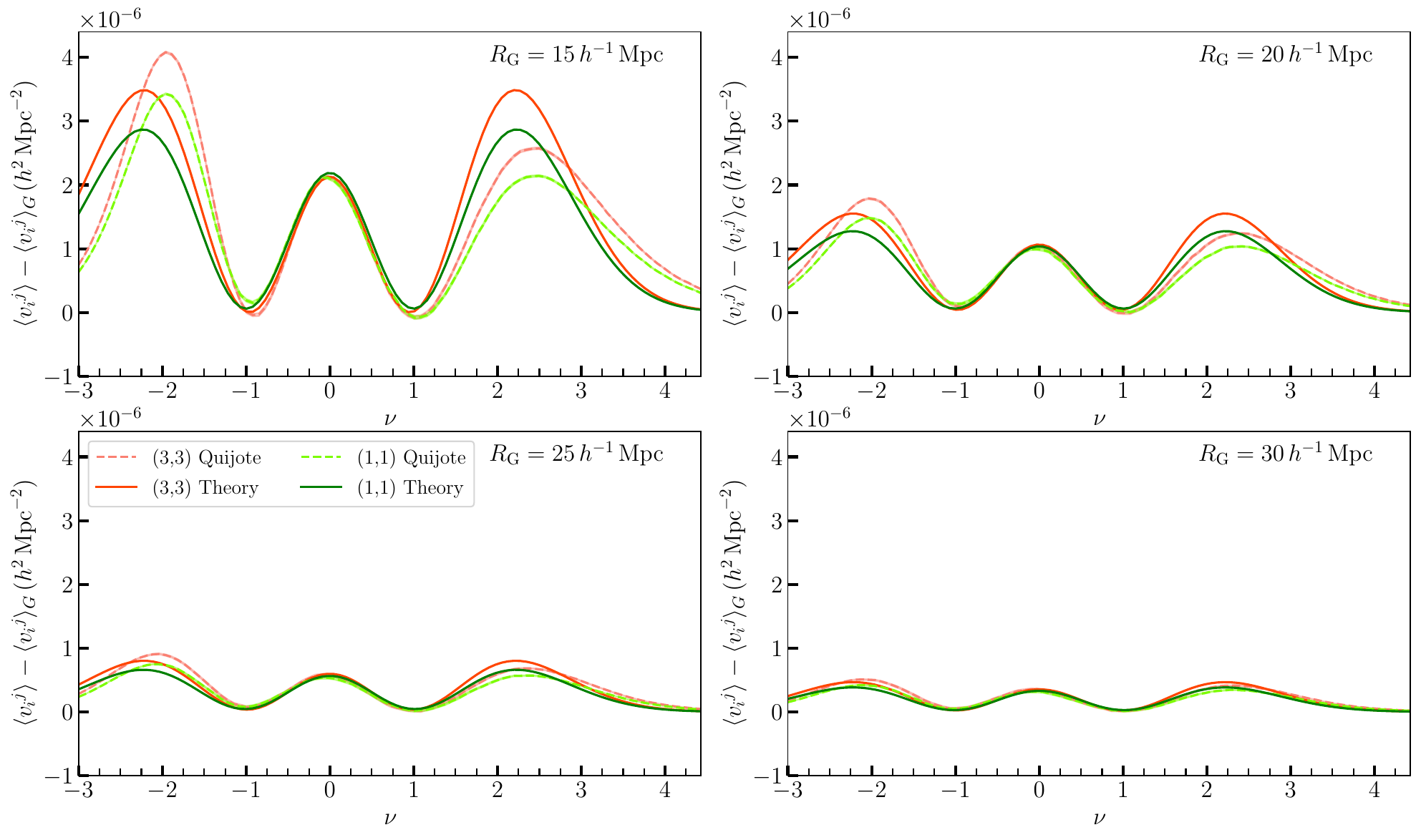}
    \caption{Residuals  $W_{2}^{0,2}|_{i}{}^{i} - \langle v_{i}{}^{i} \rangle_\mathrm{G}$ (red/green dashed lines) and $\langle v_{i}{}^{i}\rangle - \langle v_{i}{}^{i} \rangle_\mathrm{G}$ (red/green solid lines), where $W_{2}^{0,2}|_{i}{}^{i}$ is the mean value of the Minkowski tensor components from the snapshot boxes, $\langle v_{i}{}^{i}\rangle$ is the non-Gaussian prediction obtained in this work and $\langle v_{i}{}^{i} \rangle_\mathrm{G}$ the Gaussian limit of the prediction. The residuals are dominated by the even Hermite polynomials $H_{0}(\nu)$, $H_{2}(\nu)$ and $H_{4}(\nu)$. 
    Qualitatively, the same convergence is observed as in Figure~\ref{fig:res1} for $W_{1}^{0,2}|_{i}{}^{i} $.}
    \label{fig:res2}
\end{figure}

In Figures \ref{fig:1c} and \ref{fig:res2}, we present the same results as Figures \ref{fig:1b} and \ref{fig:res1} respectively, but for the second Minkowski tensor $W^{0,2}_{2}$ and the corresponding ensemble averages in Section \ref{sec:ens_v}. We observe a very similar picture; perturbation theory performs well for $R_\mathrm{G} \geq 20 \, h^{-1} \, {\rm Mpc}$, but higher order contributions increasingly impact the components on smaller scales. The residuals in Figure \ref{fig:res2} are now dominated by the even Hermite polynomials $H_{0}(\nu)$, $H_{2}(\nu)$, $H_{4}(\nu)$. 

The Figures illustrates the success for perturbation theory and the Edgeworth expansion, showing a good match between the ensemble average and measurements from simulations. In contrast, in Figures \ref{fig:app1} and \ref{fig:app2} we present the same measurements, but now the theoretical predictions for $\langle w_{i}{}^{i}\rangle$ and $\langle v_{i}{}^{i} \rangle$ and their Gaussian limits (cf. black dashed, grey dotted lines) are constructed by setting $\sigma_{v} = \sigma_{B} = 0$. Now, we see that perturbation theory performs significantly worse for the $(3,3)$ component of the statistics (cf red points/error bars), except when smoothing on the largest scales $R_\mathrm{G} = 30 \, h^{-1} \, {\rm Mpc}$. In contrast, the components perpendicular to the line of sight (green points) remain well described by perturbation theory and are not strongly affected by the Finger of God effect. It is clear that the components of the field parallel to the line of sight are significantly contaminated by non-linear velocity dispersion effects, a result also noted in \cite{Appleby:2022itn}. Whether this is positive or negative depends on whether we are interested in modeling the Finger of God effect, or if we consider it a contaminant to cosmological parameter estimation. Also, we note that the FoG effect will be more pronounced in dark matter fields compared to galaxies: dark matter particles generally exhibit larger stochastic velocities. Hence this work can be considered as a worst case scenario. In the future we will extract the Minkowski tensors from galaxy fields and consider the effect of galaxy bias, shot noise and Finger of God velocity dispersion. 

In a number of previous works, it was proposed that redshift space distortion information can be extracted from the scalar Minkowski Functionals by creating two-dimensional slices of galaxy data in planes, rotating the direction of the plane relative to the line of sight and measuring variations in the two-dimensional scalar statistics \citep{Matsubara:1995wj,Codis:2013exa}. While this approach would certainly yield directional information, the Minkowski Tensors represent a more comprehensive and natural method for extracting anisotropic signals. There are two primary advantages to using the Minkowski Tensors. First, the tensors contain additional information in their off-diagonal elements and if the density field exhibits anisotropy beyond the known redshift space distortion correction, it will present as non-zero off-diagonal elements. This information would not be present in the scalar functionals.

Second, the actual redshift space distortion effect is radial and not plane-parallel. This is particularly important for modern galaxy surveys, which cover a significant fraction of the sky. We cannot construct a plane that is globally parallel to the line of sight, but the issue is naturally resolved by using the Minkowski tensors in a spherical coordinate system, as detailed in \cite{Appleby:2022itn}. The Cartesian components constructed in this work can be simply adapted to a spherical coordinate system, by defining a coordinate system at each point in the space that is locally aligned to the line of sight and then averaging over the entire volume.

\section{Discussion}
\label{sec:disc}

Minkowski Tensors are a geometrical tool for analyzing the anisotropic information embedded in cosmic structures, especially those affected by redshift space distortions. By incorporating a perturbative non-Gaussian Edgeworth expansion (up to cubic order cumulants of the redshift space density field and its derivatives), the derived MTs can also capture the morphological evolution of the matter distribution in the weakly non-linear regime, hence improving constraints on cosmological parameters. This provides a pathway for using ongoing (DESI, Euclid) and upcoming (LSST, Roman) galaxy surveys  to infer properties of the underlying cosmology and its cosmic evolution.

In this work we have calculated the ensemble average of the translation invariant Minkowski tensors $W_{1}^{0,2}$ and $W_{2}^{0,2}$ in redshift space to leading order in a non-Gaussian expansion of the PDF of the field and its first and second derivatives. Starting from the definitions of $W_{1}^{0,2}$ and $W_{2}^{0,2}$, we utilised the assumed statistical homogeneity of the fields to approximately equate volume and ensemble averages. The ensemble average was then constructed using the Edgeworth expansion of the PDF of the fields, truncated at the level of the three point cumulants. We estimated the two- and three-point cumulants using standard perturbation theory in redshift space. 

The main results of this work are equations (\ref{eq:w1},
\ref{eq:v11}),
which express the ensemble averages of the Minkowski tensor components in terms of Hermite polynomials,  a set of coefficients which are functions of $\lambda= \sigma_{1\parallel}/\sigma_{1\perp}$ only, and two- and three-point field cumulants. To calculate the analytic predictions for a given cosmology, one should construct the linear matter power spectrum $P_{m}(k,z)$ using e.g. CAMB, perform the two-point (equations~\ref{eq:ss})
 and three-point (equations~\ref{eq:x3}) cumulant integrals, and use them in our expressions for the ensemble averages $\langle w_{i}{}^{j}\rangle$, $\langle v_{i}{}^{j}\rangle$.

We tested the analytic expressions for $\langle w_{i}{}^{j}\rangle$ and $\langle v_{i}{}^{j} \rangle$ by extracting the Minkowski tensors from $z=1$ dark matter snapshot boxes using the quijote simulation suite. We find reasonable agreement between the measured volume averages and predicted ensemble averages when we smooth the field on scales $R_\mathrm{G} > 20 \, h^{-1} \, {\rm Mpc}$, although we must include a parameterization to account for the Finger-of-God effect. This latter phenomena significantly impacts the components of the tensor parallel to the line of sight, and requires an additional free parameter to model. This will partially confound our ultimate goal of constraining cosmological parameters. 

By moving beyond the Gaussian limit, the information content of the MT curves is manifested. The Gaussian information is contained in the amplitudes of $W^{0,2}_{1}$ and $W_{2}^{0,2}$ (coefficients of $H_{0}$ and $H_{1}$ respectively), which are functions of the ratio $\sigma_{1\perp}/\sigma$ and coefficients involving $\lambda= \sigma_{1\parallel}/\sigma_{1\perp}$, which is also a ratio. Hence they contain information relating to the shape of the power spectrum: predominantly the parameters $\Omega_{m}$ and $n_{s}$, and also the growth rate $\beta = f/b$. Specifically, the amplitude of the density fluctuations $b_{1}\sigma_{8}$ cannot be extracted from the Gaussian part. However, by measuring the skewness of the MT curves, the cubic cumulants $\langle x^{3} \rangle /\sigma^{3}$ etc are proportional to $b_{1}\sigma_{8}$, so the combined measurement of the Gaussian and non-Gaussian information will provide joint constraints on $f\sigma_{8}$ and $b_{1}\sigma_{8}$.

The statistics considered in this work are tensors, and we utilized Cartesian coordinates aligned with the line of sight between observer and the density field and further assumed the plane parallel approximation. The latter condition makes the simplifying assumption that the line of sight vector is parallel at every point in the density field. In reality the redshift space distortion operator is radial relative to the observer position, and great care should be taken with coordinate systems when extracting these statistics from galaxy data. This issue was the focus of the work of \cite{Appleby:2022itn}, with the conclusion that spherical coordinate systems should be used when measuring the Minkowski tensors. In the Cartesian basis used in this work, the tensors are diagonal: they would also be diagonal in any rotated coordinate system that preserves the line of sight direction due to the statistical isotropy in the plane perpendicular to ${\bf e}_{3}$. We could potentially use the invariance of the Minkowski tensors to this subset of rotations to test statistical isotropy in the $({\bf e}_{1},{\bf e}_{2})$ plane. However, testing isotropy perpendicular to the line of sight is a subtle problem due to the fact that the subspace is $S^{2}$ not $\mathbb{R}^2$, and will be considered elsewhere.

Throughout this work we have focused on the rank-two quantities $W_{1}^{0,2}$ and $W_{2}^{0,2}$. An interesting future direction would be to study the higher-rank statistics involving increasing tensor products of $\hat{n} \otimes \hat{n} \otimes \hat{n} \otimes \dots$ . The higher rank tensors could be similarly Edgeworth expanded, and would yield different combinations of cumulants and Hermite polynomials whilst simultaneously probing a higher multipole expansion of the random shapes generated by iso-field surfaces of random fields. 

The Minkowski functionals and tensors contain complementary information relative to two-point function analyses (correlation function and power spectrum), and can be measured in conjunction with those statistics to improve existing parameter constraints. Although the Edgeworth expansion has only been utilized to cubic order in redshift space \citep{Codis:2013exa} (and quartic in real space \citep{Gay:2011wz,Matsubara:2020knr}), we can proceed to higher orders with relative ease, limited only by our ability to express the higher point cumulants using perturbation theory and performing the higher dimensional integrations necessary to obtain the $n$-point cumulants $\langle x^{n} \rangle,$ etc. 

This work serves as a precursor to extracting the Minkowski tensors from current and future galaxy catalogs, and subsequently performing cosmological parameter estimation. Principally our goal will be to infer $f\sigma_{8}$ and $b\sigma_{8}$, and by making repeated measurements at different redshifts,  potentially place constraints on the growth rate \citep{Gay:2011wz}. 
Note that this inference will undoubtedly 
have different biases compared to alternative methods such as weak lensing \citep{yamamoto2025darkenergysurveyyear} or baryonic acoustic oscillations  \citep{desicollaboration2025desidr2resultsii}.
Hence MT should be implemented to leverage such biases.
The application of our results to data will be pursued in future work.

\bibliography{refs}{}
\bibliographystyle{aasjournalv7}

\begin{acknowledgements}
SA is supported by an appointment to the JRG Program at the APCTP through the Science and Technology Promotion Fund and Lottery Fund of the Korean Government, and was also supported by the Korean Local Governments in Gyeongsangbuk-do Province and Pohang City. 
CP thanks Katarina Kraljic for feedback.
CP is partially supported by the grant \href{https://www.secular-evolution.org}{\emph{SEGAL}} ANR-19-CE31-0017
of the French Agence Nationale de la Recherche. CBP is supported by KIAS Individual Grants (PG016903) at the Korea Institute for Advanced Study, and is also supported by the National Research Foundation of Korea (NRF) grant funded by the Korean government (MSIT; RS-2024-00360385).
\end{acknowledgements}

\appendix

\section{Isotropic Limit}

In the main body of the text we constructed the ensemble averages of the Minkowski Tensors in redshift space. Here we show that in the isotropic limit, the trace of these quantities reduces to $W_{1}$ and $W_{2}$ presented in \cite{2003ApJ...584....1M}. The isotropic limit corresponds to taking $\lambda \to (1/\sqrt{2})^{+}$, $\sigma_{v} = \sigma_{B} = 0$ and the cumulants take values -- $\sigma^{2} = \sigma_{0}^{2}$, $\sigma_{1\parallel}^{2} = \sigma_{1}^{2}/3$, $\sigma_{1\perp}^{2} = 2\sigma_{1}^{2}/3$, $\sigma_{2\perp}^{2} = 8\sigma_{2}^{2}/15$, $\sigma_{2\parallel}^{2} = \sigma_{2}^{2}/5$, $\sigma_{Q}^{2} = 2\sigma_{2}^{2}/15$. 
We also have the following three-point isotropic limits
\begin{equation} \langle x^{3} \rangle = {\langle \delta^{3} \rangle \over \sigma_{0}^{3}}\,, \quad
 \langle x q_{\perp}^{2} \rangle = \langle x x_{3}^{2} \rangle = {\langle \delta |\nabla \delta|^{2} \rangle \over \sigma_{0} \sigma_{1}^{2}}\,, \quad
{\langle q_{\perp}^{2} J_{1\perp} \rangle \over \gamma_{\perp}} = {3 \langle |\nabla\delta|^{2} \nabla^{2} \delta \rangle \sigma_{0} \over 4\sigma_{1}^{4}} \,,\quad
 {\langle x_{3}^{2} J_{1\perp} \rangle \over \gamma_{\perp}} = {3 \langle |\nabla\delta|^{2} \nabla^{2} \delta \rangle \sigma_{0} \over 2\sigma_{1}^{4}}\,.
\end{equation}
Taking $\lambda \to (1/\sqrt{2})^{+}$, we have 
\begin{subequations}
\begin{eqnarray}
\nonumber  & & A_{G\perp}^{(1)} \to 2\sqrt{2}/3 ,  \quad B_{\perp}^{(1)} \to 6\sqrt{2}/15 , \quad  C_{\perp}^{(1)} \to -\sqrt{2}/15 , \quad A_{G\parallel}^{(1)} \to \sqrt{2}/3 , \quad B_{\parallel}^{(1)} \to -\sqrt{2}/15 , \quad C_{\parallel}^{(1)} \to 7\sqrt{2}/30 , \\ 
\nonumber & & A_{G\perp}^{(2)} \to 4/3 , \quad B_{\perp}^{(2)} \to 16/15 , \quad C_{\perp}^{(2)} \to 4/15 , \quad D_{\perp}^{(2)} \to -12/15 , \quad E_{\perp}^{(2)} \to -4/15 , \\
\nonumber & & A_{G\parallel}^{(2)} \to 2/3 , \quad B_{\parallel}^{(2)} \to 4/15 , \quad C_{\parallel}^{(2)} \to 6/15 , \quad D_{\parallel}^{(2)} \to 2/15 , \quad E_{\parallel}^{(2)} \to -6/15 \,.
\end{eqnarray} 
\end{subequations}
Inserting these into equations (\ref{eq:w1},\ref{eq:v11}) we have 
\begin{subequations}
\begin{eqnarray}  
\langle w_{1}{}^{1} \rangle + \langle w_{2}{}^{2} \rangle &=& {2 \sigma_{1} \over 9\sqrt{3}\pi \sigma_{0}}e^{-\nu^{2}/2} \left[ H_{0}(\nu) + {1 \over 6}{\langle \delta^{3} \rangle \over \sigma_{0}^{3}}H_{3}(\nu) + {1 \over 2} {\langle \delta |\nabla \delta|^{2} \rangle \over \sigma_{0} \sigma_{1}^{2}}H_{2}(\nu) + {\cal O}(\sigma_{0}^{2}) \right]\,, \\
 \langle w_{3}{}^{3} \rangle  &=& { \sigma_{1} \over 9\sqrt{3}\pi \sigma_{0}}e^{-\nu^{2}/2} \left[ H_{0}(\nu) + {1 \over 6}{\langle \delta^{3} \rangle \over \sigma_{0}^{3}}H_{3}(\nu) + {1 \over 2} {\langle \delta |\nabla \delta|^{2} \rangle \over \sigma_{0} \sigma_{1}^{2}}H_{2}(\nu) + {\cal O}(\sigma_{0}^{2}) \right]\,,
\end{eqnarray} 
\end{subequations}
and 
\begin{subequations}
\begin{eqnarray}  \langle v_{1}{}^{1} \rangle \!+\! \langle v_{2}{}^{2} \rangle\! &=& \!{2 \sigma_{1}^{2} \over 27\pi \sqrt{2\pi} \sigma_{0}^{2}}e^{-\nu^{2}/2}\! \left[ H_{1}(\nu) \!+\! {1 \over 6}{\langle \delta^{3} \rangle \over \sigma_{0}^{3}}H_{4}(\nu) + {\langle \delta |\nabla \delta|^{2} \rangle \over \sigma_{0} \sigma_{1}^{2}}H_{2}(\nu) \!-\! {3 \over 4} {\langle |\nabla \delta|^{2} \nabla^{2}\delta \rangle \sigma_{0} \over \sigma_{1}^{4}}H_{0}(\nu) \!+\! {\cal O}(\sigma_{0}^{2}) \right]\!, \\
 \langle v_{3}{}^{3} \rangle \! &=&\! {\sigma_{1}^{2} \over 27\pi \sqrt{2\pi} \sigma_{0}^{2}}e^{-\nu^{2}/2}\! \left[ H_{1}(\nu) \!+\! {1 \over 6}{\langle \delta^{3} \rangle \over \sigma_{0}^{3}}H_{4}(\nu) \!+\! {\langle \delta |\nabla \delta|^{2} \rangle \over \sigma_{0} \sigma_{1}^{2}}H_{2}(\nu) - {3 \over 4} {\langle |\nabla \delta|^{2} \nabla^{2}\delta \rangle \sigma_{0} \over \sigma_{1}^{4}}H_{0}(\nu) \!+\! {\cal O}(\sigma_{0}^{2}) \right]\!.
\end{eqnarray} 
\end{subequations}
The traces yield the Minkowski Functionals $W_{1}$ and $W_{2}$ in \cite{2003ApJ...584....1M} respectively. 

\begin{figure}
    \centering
    \includegraphics[width=0.92\textwidth]{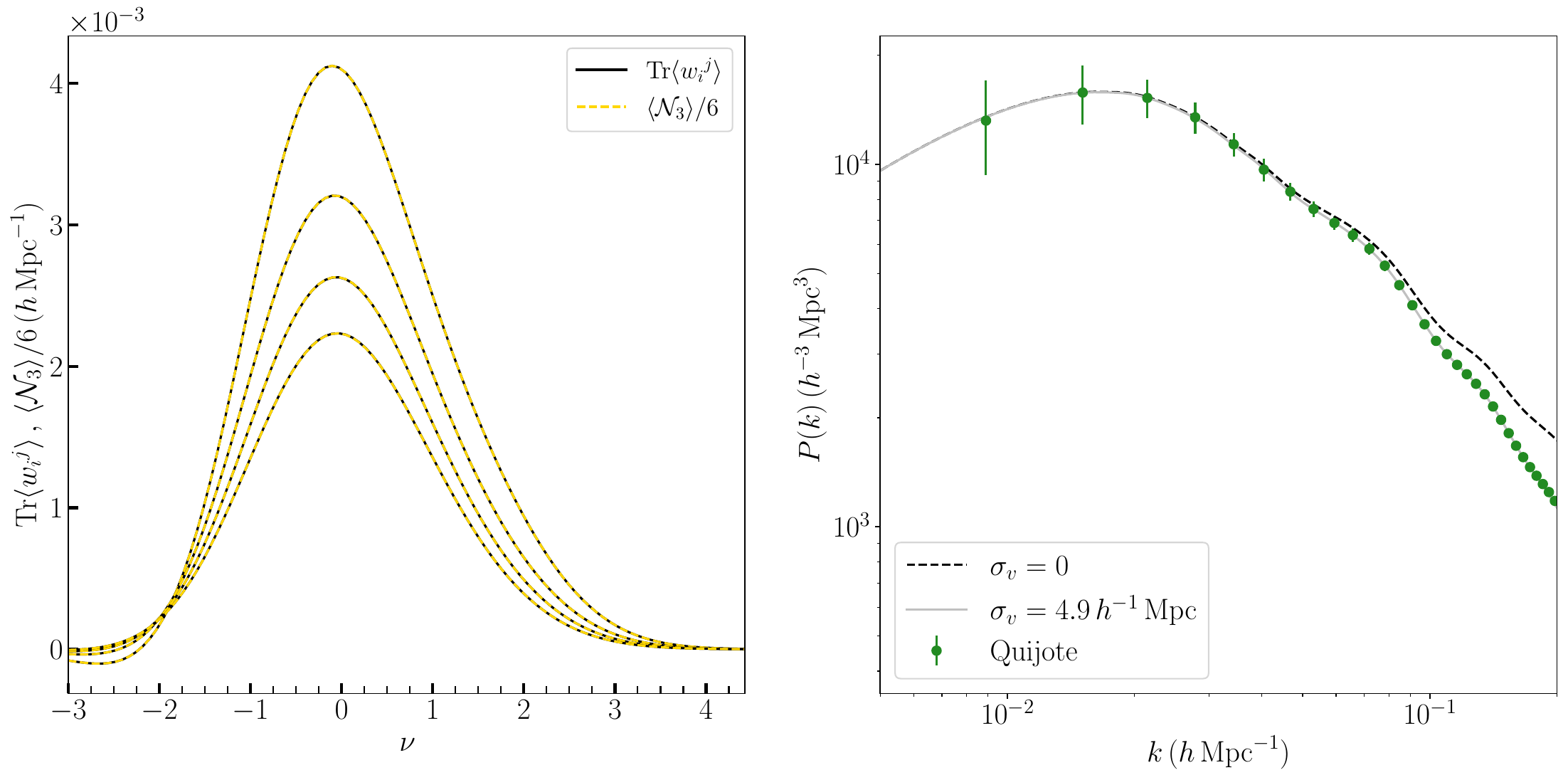}
    \caption{[Left panel] A comparison of the trace of $w_{i}{}^{j}$ calculated in this work (black solid lines) and the total area of iso-field surfaces ${\cal N}_{3}$ in \cite{Codis:2013exa} (yellow dashed lines). The curves in descending amplitude are for smoothing scales $R_\mathrm{G} = 15, 20, 25, 30 \, h^{-1} \, {\rm Mpc}$. [Right panel] The redshift space power spectrum of $z=1$ fiducial cosmology quijote snapshot boxes (green points/error bars) and the perturbation theory expectation value with zero velocity dispersion (black dashed line) and with best fit $\sigma_{v} = 4.9 \, h^{-1} \, {\rm Mpc}$ (grey solid line).}
    \label{fig:app0}
\end{figure}

The ensemble average of the Minkowski Functional $W_{1}$ has been constructed in redshift space in \cite{Codis:2013exa}; we can also compare the trace ${\rm Tr}\langle w_{i}{}^{j} \rangle$ obtained in this work to that result; specifically ${\cal N}_{3}$ as defined in equation (47, 48) of \cite{Codis:2013exa}. To make the comparison, we fix $\sigma_{v} = \sigma_{B} = 0$ since the Finger of God was not included in that work, and we divide ${\cal N}_{3}$ by $6$ to match definitions. The resulting ensemble averages of ${\rm Tr}\langle w_{i}{}^{j} \rangle$ and ${\cal N}_{3}/6$ are presented in the left panel of Figure \ref{fig:app0}; the four sets of curves correspond to $R_{\rm G} = 15, 20, 25, 30 \, h^{-1} \, {\rm Mpc}$. The black solid and yellow dashed lines representing the statistics are in excellent agreement, which serves as a consistency check of our calculation. 

\section{Finger of God Velocity Dispersion}
\label{sec:app_b}

In the main body of the text, we use velocity dispersion values $\sigma_{v} = \sigma_{B} = 4.9 \, h^{-1} \, {\rm Mpc}$. In this section we briefly review how this value has been determined. In the literature the parameters $\sigma_{v}$, $\sigma_{B}$ are sometimes treated as independent parameters to be jointly constrained, but in this work we always fix them to be equal. We also note that the choice of exponential function to describe the FoG effect is not necessarily applicable over all scales, and in fact on smaller scales the exponential damping is too steep. Improved modeling of this phenomena on small scales will be considered elsewhere. 

To estimate $\sigma_{v}$, we take $N_{\rm real} = 4000$ measurements of the dark matter power spectrum (monopole) extracted from the $z=1$, redshift space distorted fiducial quijote simulations. The measurements are made over $N_{\rm b} = 31$ fourier mode bins from $9 \times 10^{-3} \, h \, {\rm Mpc}^{-1} < k < 0.20 \, h \, {\rm Mpc}^{-1}$. From these realisations we construct a covariance matrix for the power spectrum which we denote $\Sigma_{ij}$, where $i,j$ subscripts run over the $k$ bins. We then take a different set of $N = 500$ realisations and measure the mean power spectrum in the $31$ Fourier bins; this is our data vector which we denote as $\bar{P}_{i}$. To this power spectrum, we fit the following functional form, 
\begin{eqnarray}\label{eq:th1} & & P_{i} = {1 \over 2}\int_{-1}^{1}d\mu P(\mu, k_{i}) = {1 \over 2}\int_{-1}^{1}d\mu \left[P_{00} + \mu^{2} \left( 2P_{01} + P_{02} + P_{11} \right) + \mu^{4} \left( P_{03} + P_{04} + P_{12} + P_{13} + P_{22}/4 \right) \right] , 
\end{eqnarray} 
\noindent which is derived using the combined distribution function and Eulerian perturbation theory approach of \cite{Vlah:2012ni,Saito:2014qha,Seljak:2011tx,McDonald:2009hs,Okumura:2011pb,Okumura:2012xh}. We direct the reader to those works and \cite{Howlett:2019bky}, Appendix A,  for the exact definitions of the $P_{mn}$ contributions and corresponding kernels. We fix all cosmological parameters to their correct, fiducial values, and fit a single parameter $\sigma_{v}$ to the power spectrum by minimizing 
\begin{equation}
 \chi^{2} = [\bar{P}_{i} - P_{i}] \Sigma^{-1}_{ij} [\bar{P}_{j} - P_{j}] \,.
\end{equation}
The resulting best fit is given by $\sigma_{v} = 4.90 \pm 0.04 \, h^{-1} \, {\rm Mpc}$, which is the value that is adopted in this work. In the right panel of Figure \ref{fig:app0} we present the $z=1$, redshift space power spectrum measured from the Quijote simulations (green points/error bars), the perturbation theory predictions with $\sigma_{v} = 0$ and $\sigma_{v} = 4.90 \, h^{-1} \, {\rm Mpc}$ (black dashed, solid grey lines respectively). 

To obtain the best fit $\sigma_{v}$ velocity dispersion, we have measured the power spectrum shape on large scales under the assumption that cosmological parameters are known and fixed. In practise, this is not an approach that could be used for realistic data. Instead, one should simultaneously fit $\sigma_{v}$ and cosmological parameters using a large array of summary statistics, potentially including the power spectrum, Minkowski functionals and tensors etc. Alternatively, one could measure the galaxy correlation function on small scales to infer $\sigma_{v}$.

\begin{figure}
    \centering
    \includegraphics[width=0.98\textwidth]{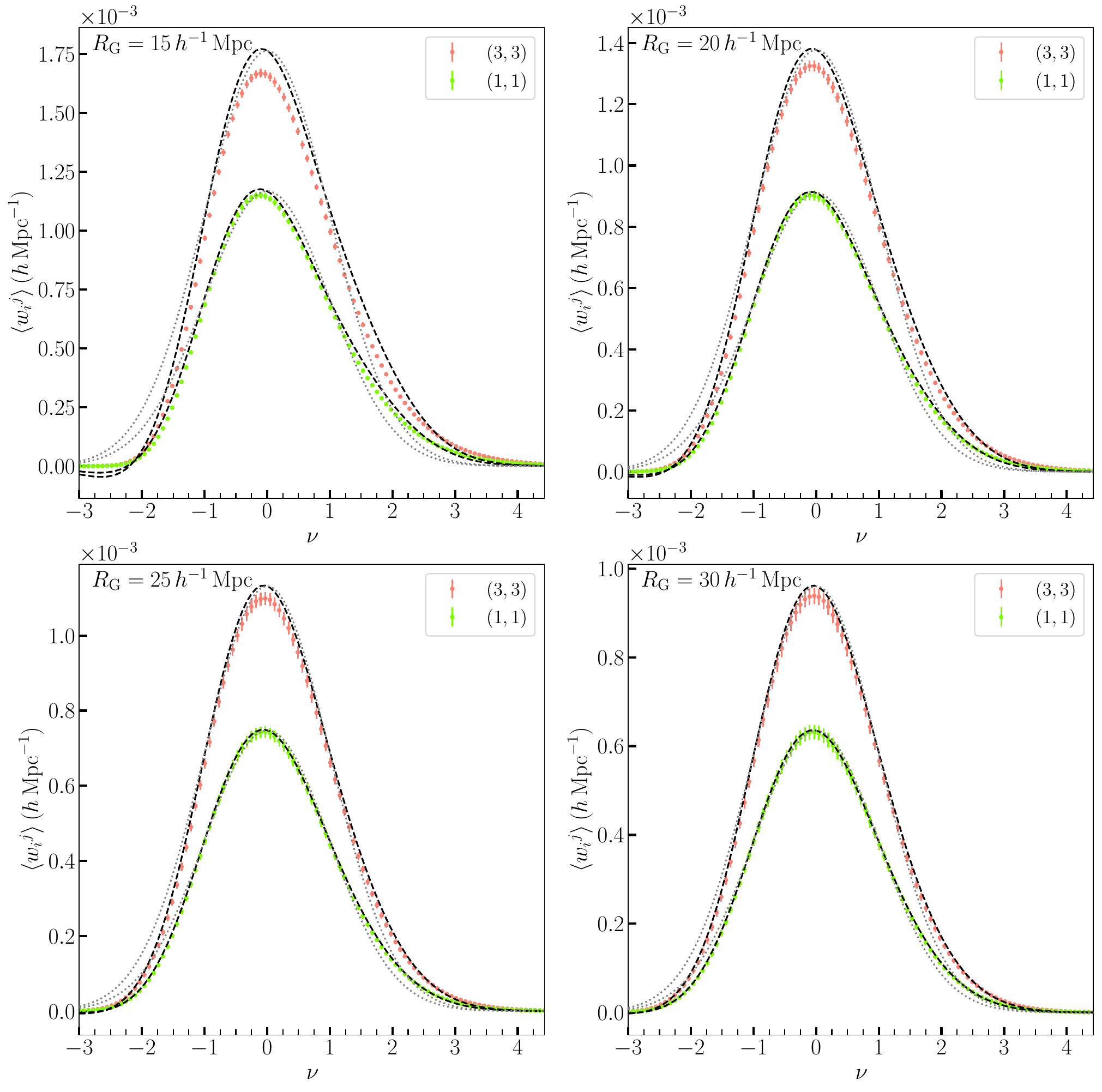}
    \caption{Same as Figure \ref{fig:1b}, but with zero FoG velocity dispersion $\sigma_{v} = \sigma_{B} = 0$ in the ensemble averages $\langle w_{i}{}^{j} \rangle$ and $\langle w_{i}{}^{j} \rangle_{G}$ (black dashed, grey dotted lines). The discrepancy along the LOS is more  significant when contrasted to Figure~\ref{fig:1b}.
    }
    \label{fig:app1}
\end{figure}

\begin{figure}
    \centering
    \includegraphics[width=0.98\textwidth]{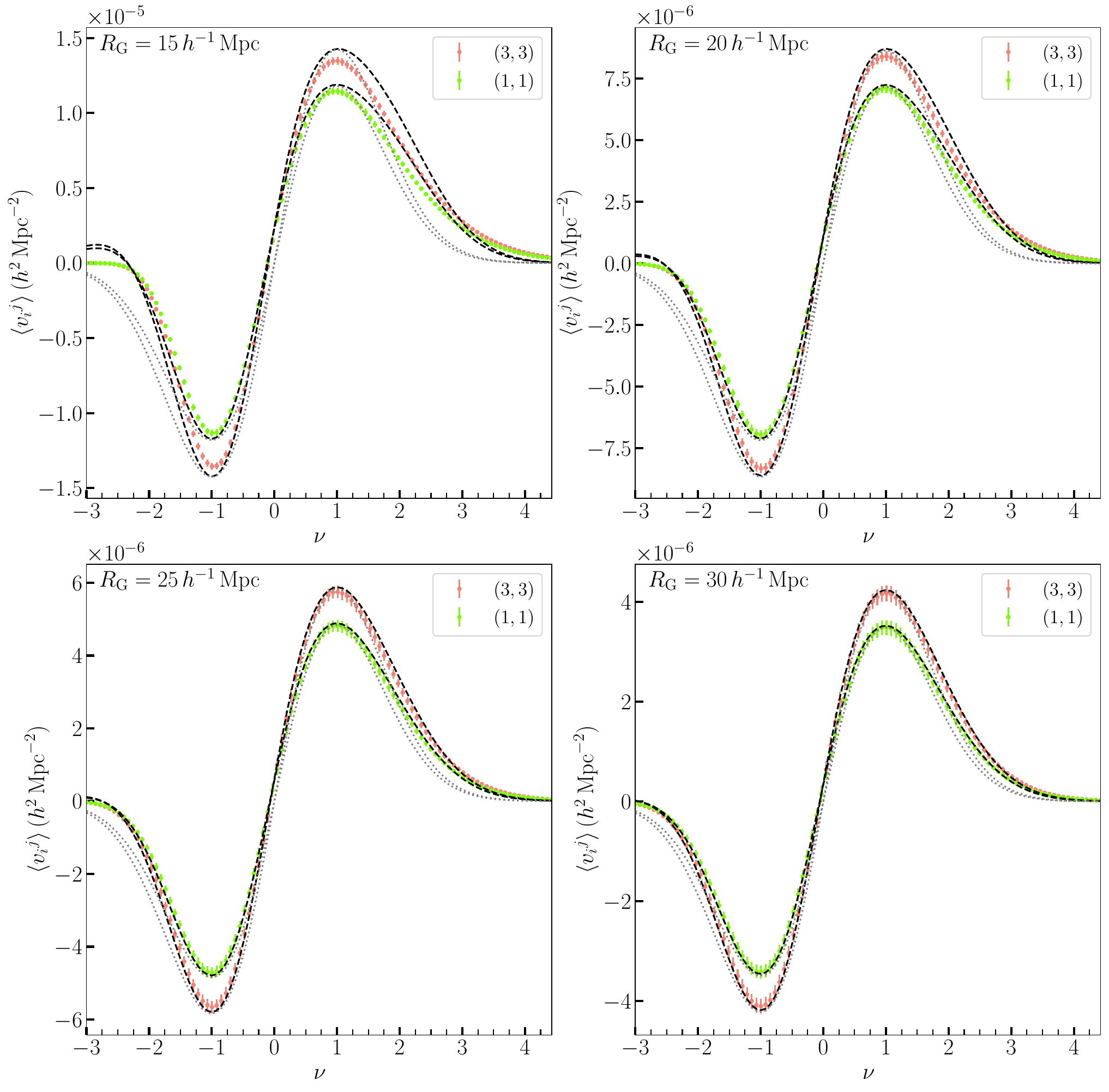}
    \caption{Same as Figure \ref{fig:1c}, but with zero FoG velocity dispersion $\sigma_{v} = \sigma_{B} = 0$ in the ensemble averages $\langle v_{i}{}^{j} \rangle$ and $\langle v_{i}{}^{j} \rangle_\mathrm{G}$ (black dashed, grey dotted lines). As in Figure~\ref{fig:app1}, 
    the discrepancy along the LOS is more  significant when contrasted to Figure~\ref{fig:1c}.
    }
    \label{fig:app2}
\end{figure}

To show the corresponding effect the Finger of God effect on the statistics used in this work, in Figures \ref{fig:app1}, \ref{fig:app2} we present the Minkowski Tensor predictions without the Finger of God effect accounted for, that is we set $\sigma_{v} = \sigma_{B} = 0$. 

\section{Finite Resolution Effects}
\label{sec:app_c}

In this work we generated discrete dark matter fields on a lattice by binning dark matter particles into a regular grid of resolution $\Delta = 1000/256 = 3.9 \, h^{-1} \, {\rm Mpc}$. In this appendix we check that our results are not sensitive to this choice. To do so, we take a small set of $N_{\rm real} = 10$ dark matter particle snapshots at $z=1$, and bin them into two different resolution grids -- $\Delta = 3.9 \, h^{-1} \, {\rm Mpc}$ and $\Delta = 1000/512 = 1.95 \, h^{-1} \, {\rm Mpc}$. The fields are then smoothed using the smallest scale considered in this work, $R_{G} = 15 \, h^{-1} \, {\rm Mpc}$ and finally we extract the Minkowski tensors $w_{i}{}^{j}$ and $v_{i}{}^{j}$. We denote the measured MTs as $\langle w_{i}{}^{j}\rangle_{256}$, $\langle v_{i}{}^{j}\rangle_{256}$ and $\langle w_{i}{}^{j}\rangle_{512}$, $\langle v_{i}{}^{j}\rangle_{512}$ from the low and high resolution fields respectively. 

\begin{figure}
    \centering
    \includegraphics[width=0.98\textwidth]{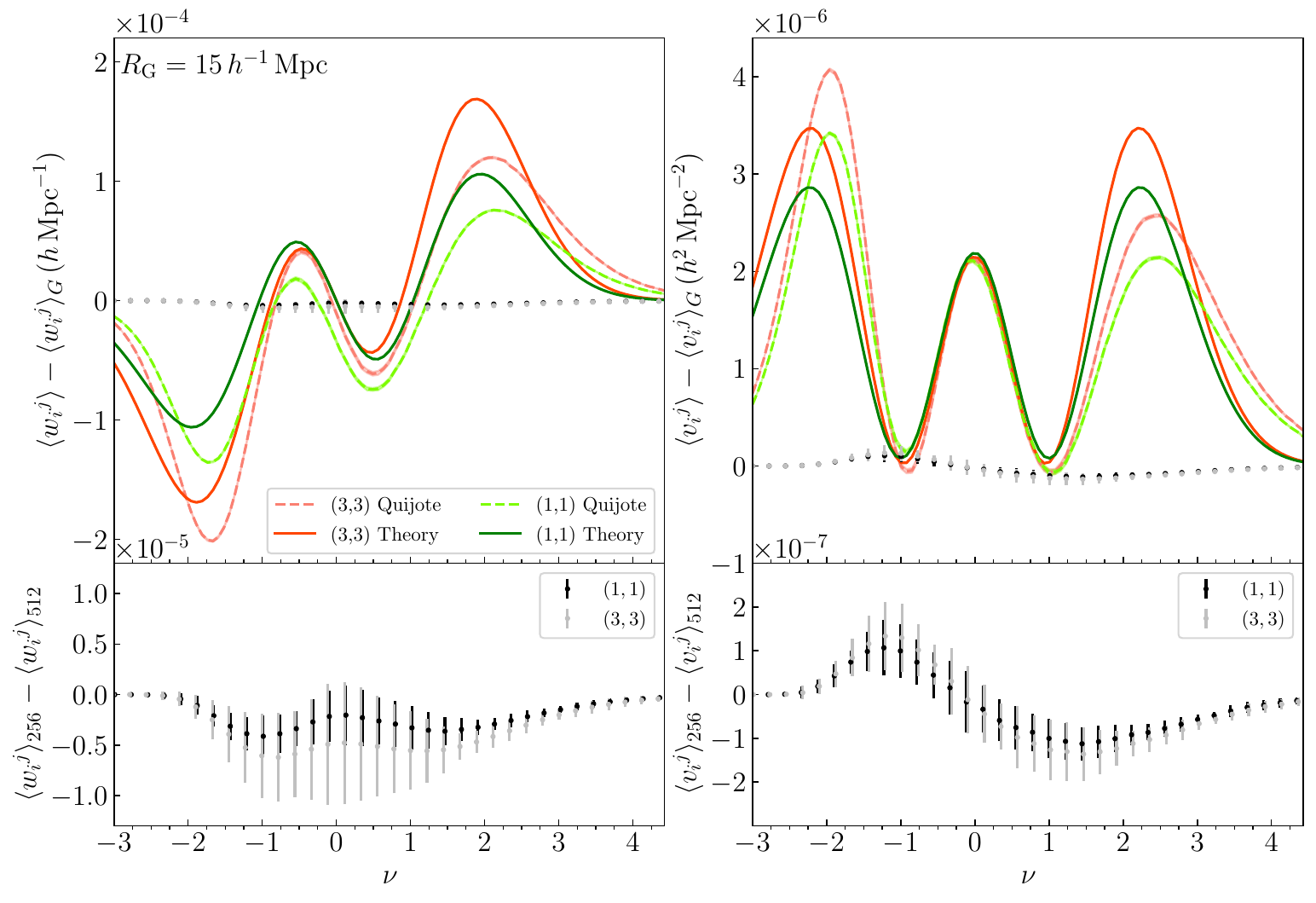}
    \caption{[Top Panels] The difference between measured values of the Minkowski tensors ($w_{i}{}^{j}$ (left) and $v_{i}{}^{j}$ (right)) and their Gaussian expectation values in redshift space, for a field smoothed using $R_{G} = 15 \, h^{-1} {\rm Mpc}$. The green/red data are reproduced from the top left panels of Figures \ref{fig:res1},\ref{fig:res2}. The black/grey points correspond to the residuals between Minkowski tensors as measured in high $(\Delta = 1.95 \, h^{-1} \, {\rm Mpc})$ and low $(\Delta = 3.9 \, h^{-1} \, {\rm Mpc})$ resolution fields. The black/grey points are the $(1,1)$ and $(3,3)$ components of the tensors respectively. [Bottom Panels] The same black/grey points as in the top panel, but using a smaller $y$-axis range.
    }
    \label{fig:app_c}
\end{figure}

In the top panels of Figure \ref{fig:app_c} we present the residuals $\langle w_{i}{}^{j}\rangle - \langle w_{i}{}^{j}\rangle_{G}$ and $\langle v_{i}{}^{j}\rangle - \langle v_{i}{}^{j}\rangle_{G}$ : these are the non-Gaussian signals that we are attempting to measure. These panels are the same as the top left panels of Figures \ref{fig:res1} and \ref{fig:res2}. In these panels we also include the difference between the Minkowski tensors $\langle w_{i}{}^{j}\rangle_{256} - \langle w_{i}{}^{j}\rangle_{512}$ and $\langle v_{i}{}^{j}\rangle_{256} - \langle v_{i}{}^{j}\rangle_{512}$ as black/grey points/errorbars. The points and errorbars are the mean/error on the mean from the $N_{\rm real} = 10$ realisations and the black/grey points correspond to $(1,1)$ and $(3,3)$ components respectively. These points are representative of the magnitude of the effect of finite resolution on our results, and are negligible relative to the signal that we are attempting to measure. In the lower panels of the Figure we only plot the $\langle w_{i}{}^{j}\rangle_{256} - \langle w_{i}{}^{j}\rangle_{512}$ and $\langle v_{i}{}^{j}\rangle_{256} - \langle v_{i}{}^{j}\rangle_{512}$ residuals using a smaller $y$-axis range. The lower panels present some non-random dependence of the residuals on $\nu$, indicating some systematic effect due to finite resolution, but the effect is an order of magnitude smaller than the non-Gaussian residuals. We note that the effect of finite resolution on the Minkowski functionals has been considered previously, with similar conclusions \citep{2005ApJ...633....1P,Kim:2014axe,Appleby:2017ahh}.

\end{document}